\documentclass[conference]{IEEEtran}
\usepackage{epsfig}
\usepackage{graphicx}
\usepackage{subfigure}
\usepackage{amssymb}
\usepackage{makeidx}
\usepackage{amsmath}
\usepackage{amsthm}
\usepackage{graphicx}
\usepackage{epsf}
\usepackage{epsfig}
\usepackage{psfig}
\usepackage{ccaption}
\usepackage{array}
\usepackage{tabularx}
\usepackage{multirow}
\usepackage{epsfig}
\usepackage{cite}
\usepackage{procedure,algorithm,algorithmic}

\begin{document}

\title{Imaging Using Millimeter Wave Communication Networks: A Bonus SAR}

\author{\IEEEauthorblockN{Husheng Li}
\IEEEauthorblockA{
The University of Tennessee, Knoxville,
TN, USA \\
hli31@utk.edu}
}

\maketitle

\begin{abstract}
In the next generations of cellular communication networks, higher density of base stations and higher frequency bands will be adopted. If being reflected by targets, the communication signal also brings information of the targets, in addition to the communication messages, to the receivers. In this paper, it is proposed to leverage the reflected communication signals to reconstruct an image of the environment. Due to the analogy to traditional synthetic aperture radar (SAR) and inverse SAR (ISAR), the principles of SAR and ISAR, namely the tomography via Fourier transformation, are adopted with necessary improvements. The algorithms are further refined to estimate the 3-dimensional silhouette of the environment. Numerical simulations are carried out to demonstrate the proposed algorithms.
\end{abstract}

\section{Introduction}

5G wireless communication networks have recently attracted considerable attentions \cite{Agiwal2016}. Future 5G communication systems are expected to support millions of devices within a local area (the mMTC mode), peak transmission rates of Gbps (the eMBB mode), as well as very low latency and ultra reliability in the radio access (the uRLLC mode). Motivated by the progress of 5G networks, the 6G networks are provisioned by researchers and industry. For simplicity, the 5G and beyond networks are called the 5G+ networks. A common knowledge in the community of wireless communications is that the environment causes substantial impact on communication signals. For example, the reflection of communication signals at various reflectors will substantially attenuate the signal strength, thus degrading the signal-to-noise ratio (SNR). In particular, for higher data transmission rate and ultra-short delay, the 5G+ networks expect to use higher frequency bands, such as the millimeter wave (mmWave) and TeraHz bands. These high frequency bands are even more susceptible to the environment, such as rain fade (e.g, 30\% SNR drop)\cite{ZhaoQ2006}, blockage (complete loss of signal by an obstacle) \cite{Singh2008}, transceiver displacement due to wind (a 2.5mm displacement of transmitter incurs a 50\% change of phase in the 60GHz band) \cite{mmWave_Li_2017_14}. However, despite being bad news to communications, the sensitivity of communication signals to the environment could benefit the task of sensing, as the other side of the same coin, since the environment information is embedded in the impact on communication signals. Therefore, the negative impact of environment on 5G+ communication signals can be leveraged to sense the environment. From an alternative perspective, sensing needs illuminations (e.g., eyes see objects by sunshine illumination; radar senses targets by proactively sending out illuminating electromagnetic (EM) waves). Then, the communication signal in 5G+ networks can serve as the illumination, similarly to a spotlight using invisible EM waves, thus enabling the capability of sensing.  

\begin{figure}
  \centering
  \includegraphics[scale=0.35]{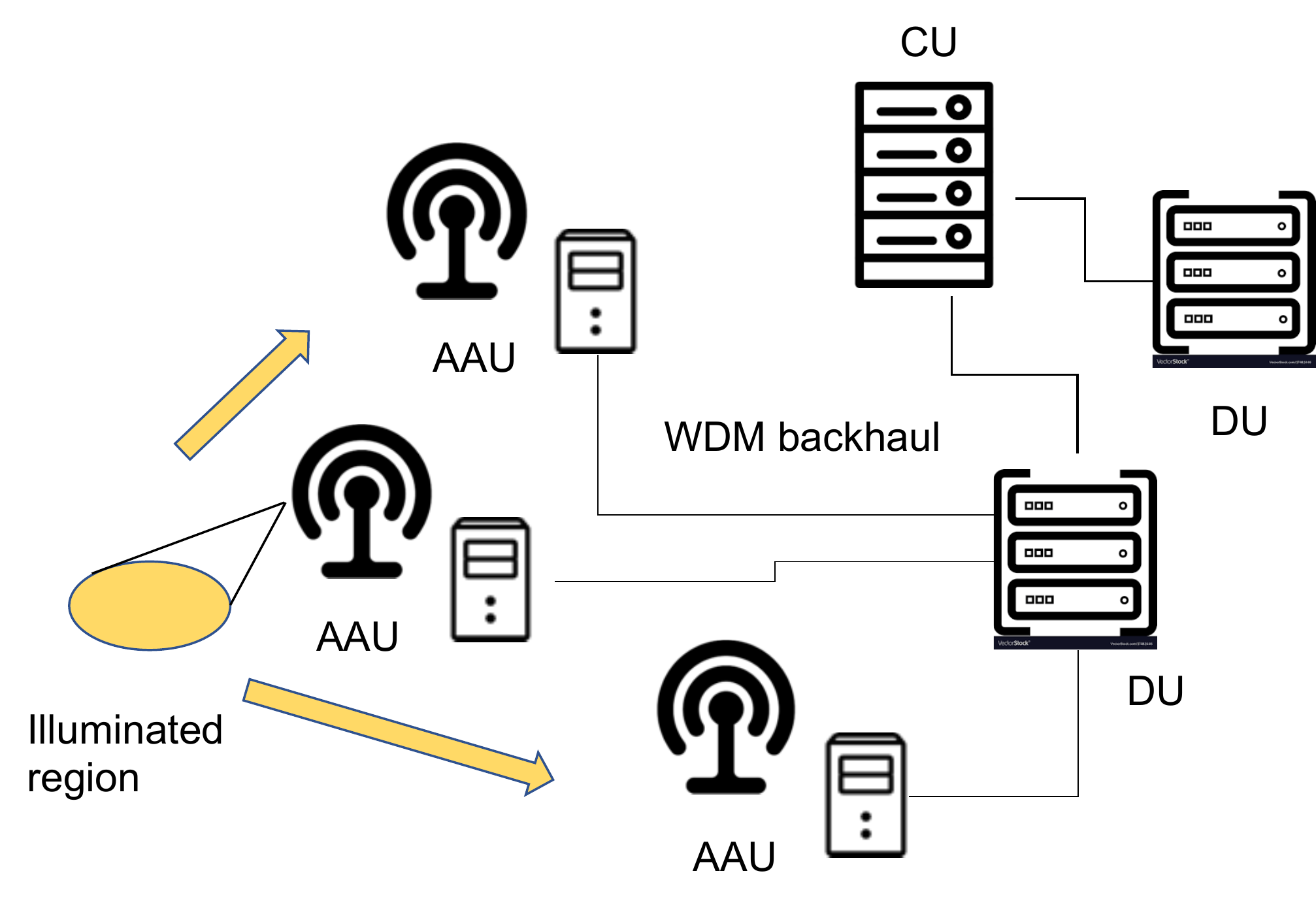}
  \caption{A possible network architecture.}\label{fig:CUDU}
\end{figure}

In this paper, we study the outdoor environment imaging using communication signals. This communication network based imaging is expected to work as follows: as illustrated in Fig. \ref{fig:CUDU}, a base station sends out communication signals, which are reflected in the environment and received by other base stations (only one is illustrated in the figure); the transmitted and received signals are both sent to a processing center in the network to reconstruct the image (or more precisely, the scattering coefficient) of the illuminated area. After many rounds of transmission and reception at different locations, a global image of the environment is expected to be reconstructed. We consider only the base stations for signal transmission and reception and the core network for data processing in this imaging procedure, since the precise positioning and measurement feedback of user equipments (UEs) are difficult. The environmental imaging will find many applications after image analysis and understanding (which is beyond the scope of this paper), such as detecting speeding vehicles (by identifying moving reflectors), monitoring flooding (by detecting water like reflectors), social distance surveillance (by identifying human-like targets), subdivision intruder detection (by locating pedestrian-like targets) and plantation monitoring (by identifying the corresponding texture of plants). 

The remainder of this paper is organized as follows. In Section \ref{sec:related}, the related works are briefly introduced. The system model is explained in Section \ref{sec:model}. The principles of SAR and ISAR systems are introduced in Sections \ref{sec:principles}. Then, the imaging algorithms based on the principles of SAR and ISAR are detailed in Sections \ref{sec:SAR} and \ref{sec:ISAR}, respectively. The system performance is analyzed theoretically and numerically in Sections \ref{sec:performance} and \ref{sec:numerical}, respectively. The final conclusions are given in Section \ref{sec:conclusion}.

\section{Related Work}\label{sec:related}
The proposed imaging procedure can be applied to the detection and inference of human or other (e.g., vehicle) targets.  It can be considered as a multi-static radar network (i.e., the transmitter and receiver are located at different positions) using communication signals for illumination, or a video camera network in the microwave or mmWave band. 
\begin{itemize}
\item Comparison with multi-static radar system: Although the proposed sensing scheme leverages many approaches in radar systems, the infrastructure is not dedicated to radar sensing, thus bringing novel challenges and solutions. 

\item Comparison with Video Camera Network: The image resolution of the proposed communication signal based sensing is expected to be much coarser, due to the much larger wavelength and much less antennas (as pixels). This disadvantage happens to be a merit in terms of privacy protection, since the image resolution can hardly infer the identifies of people. Meanwhile, the communication network itself is the source of illumination, which makes the proposed scheme work at any time. In a contrast, the video camera network is inoperable at night unless there is a light illumination. Finally the video camera network needs extra cost such as towers and communications, while the proposed scheme leverages existing infrastructure with little extra expenditure. 
\end{itemize}

There have been excellent studies on leveraging communication signals for imaging \cite{Zhu2015,Zhu2015_2,Wu2020,Arroyo2013}, particularly in the mmWave band. In a contrast, these studies are mainly focused on indoor imaging for the purpose of inferring locations and surface properties of targets, based on monostatic transceivers, while the our proposed network-wide sensing is mainly for \textit{outdoor} imaging using wide-area cellular network. Moreover, in \cite{Zhu2015,Zhu2015_2} only the information of RSS and angle of arrival (AOA) are used, which limits the imaging to targets having strong directionality of reflections. In \cite{Wu2020}, full-duplex Qualcomm IEEE 802.11ad/ay chipsets in the 60GHz band are used for multi-person locationing by leveraging periodic pulses and beamforming. The corresponding imaging is based on estimating the silhouette by radar ranging, instead of estimating the scattering coefficients of the target. So are the studies in \cite{Aladsani2019,Sturn2007}. The studies on OFDM signal based passive radar \cite{Arroyo2013,Braun2010} are closer to the PIs' proposed imaging scheme, since both consider wide-band OFDM signals which are used in 4G and 5G cellular systems. However, these studies detect only significant scatterers; e.g., in the experiment of \cite{Arroyo2013}, only a few artificially placed reflectors are located while other less significant targets are omitted. They are also based on ranging using a single pair of transmitter and receiver, while the proposed scheme is rooted from the Fourier transform framework and leverages a wide-area network for a collaborative imaging.

\section{System Model}\label{sec:model}
We assume that there are totally $N$ base stations within a certain area, with perfect time synchronization. Each could be transmitting or receiving. When transmitting, a base station uses a frequency band with $M$ subcarriers; the carrier frequency of the first subcarrier is $f_c$, while the frequency gap between adjacent subcarriers is $\delta f$. Each base station is equipped with $N_a$ antennas, which could be arrayed in one dimension or two dimensions. The height of each base station is assumed to be perfectly known.

We assume that the ground is flat, which will be relaxed later when we discuss 3-dimensional reconstruction. When a base station transmits and its signal is reflected by the ground, a region of the ground is illuminated. The reflectivity (or scattering coefficient) at position $(x,y)$ is a complex number, which can be written as
\begin{eqnarray}
g(x,y)=|g(x,y)|e^{j\phi(x,y)}.
\end{eqnarray}
Note that the imaginary part of the reflectivity stems from random phase change at the reflectors. In practice, the phase part $e^{j\phi(x,y)}$ changes much more radically than the magnitude $|g(x,y)|$, which is of key importance in the image reconstruction when only a portion of the frequency spectrum can be measured. This will be detailed later.

It is assumed that the base stations send their raw measurements to a data processing center, which endeavors to reconstruct the image of illuminated regions. A possible network architecture is illustrated in Fig. \ref{fig:CUDU}: the active antenna units (AAUs) transmit or receive radio frequency (RF) signals, while the measurements can be collected at the distributed units (DUs) and further centralized units (CUs), via WDM optical fibers, for further data processing (e.g., a DU could be managing more than 30 AAUs, while a CU can control more). Note that we do not consider the measurements at mobile stations, since it could be difficult to feed back the data via the wireless link. However, it is still possible to send back less intensive data, if algorithms can be found, in which only intermediate results, instead of the raw data, need to be communicated. This will be our future study.

\section{Principles of SAR and ISAR}\label{sec:principles}
In this section, we introduce the underlying principles of SAR and ISAR systems, which will be modified for the context of communication network based sensing. 

\subsection{Principle of Spotlight SAR}\label{sec:principle_SAR}

\begin{figure}
  \centering
  \includegraphics[scale=0.3]{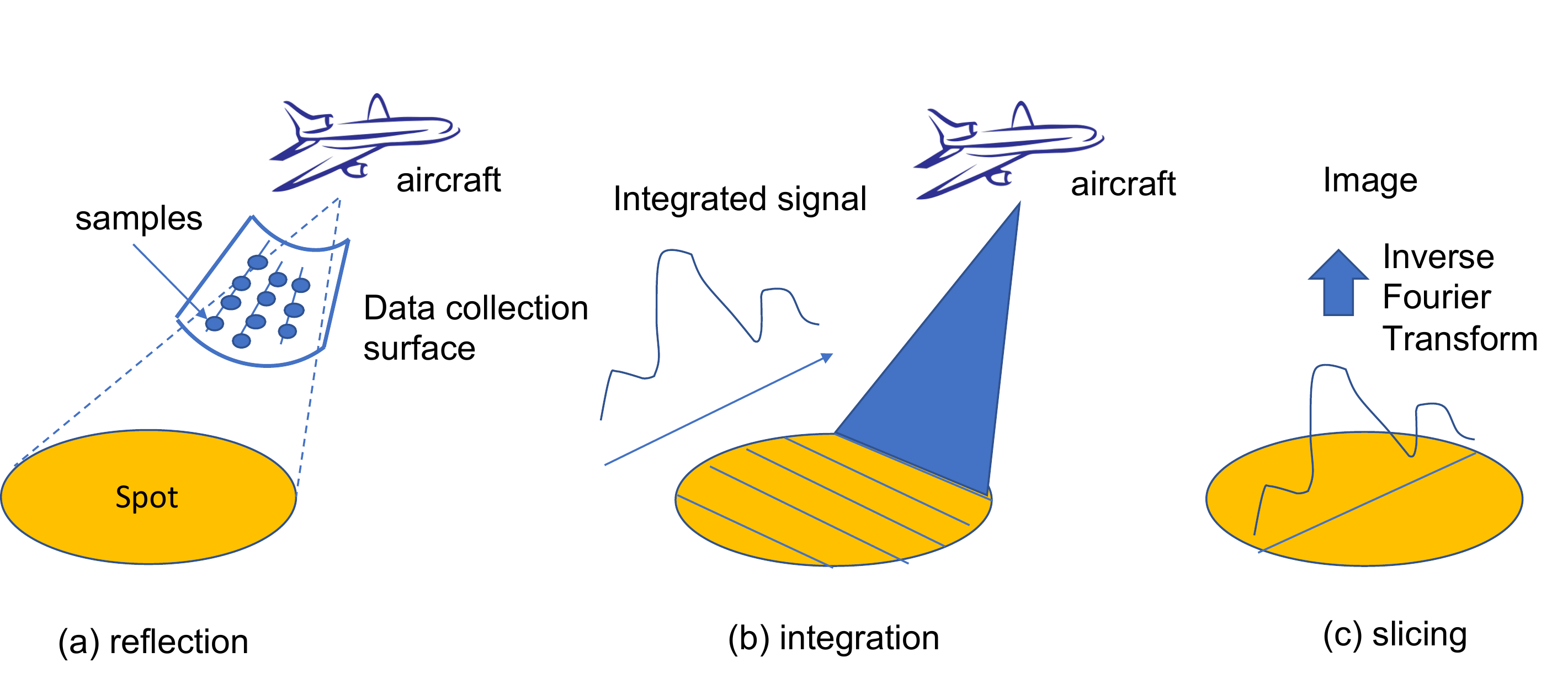}
  \caption{An illustration of spotlight SAR.}\label{fig:spotlight}
\end{figure}

The spotlight SAR imaging \cite{Jakowatz1999} is based on that of tomography, namely the projection-slice theorem. Take a 2-dimensional target, whose support region is $\Omega$, for instance. The pixels (say, the reflectivity of the surface) are given by $g(x,y)$, $(x,y)\in \Omega$. Then, the corresponding 2-dimensional Fourier transform is given by
\begin{eqnarray}
G(jX,jY)=\int_{\Omega}g(x,y)e^{-j(xX+yY)}dxdy.
\end{eqnarray}
Consider the image resulting from the projection of $g(x,y)$ to a line, say the $y$-axis. The integrated signal on the $y$-axis is given by 
\begin{eqnarray}
h(y)=\int_{\Omega} g(x,y)dx,
\end{eqnarray}
whose Fourier transform is denoted by 
\begin{eqnarray}
H(jY)=\int_{\Omega} h(y)e^{-jyY}dy. 
\end{eqnarray}
Then, the projection-slice theorem states that the Fourier transform $H(jY)$ equals the values of $G(jX,jY)$ on the line passing the origin and being parallel to the $y$-axis, namely
\begin{eqnarray}
H(jY)=G(0,jY). 
\end{eqnarray}
Based on the projection-slice theorem, the mechanism of spotlight SAR imaging is illustrated in Fig. \ref{fig:spotlight}. The aircraft illuminates the given region by sending out radar pulses, as shown in Fig. \ref{fig:spotlight}. Then, the radar signal reflected by the targets on the line orthogonal to the illuminating beam (as shown in Fig. \ref{fig:spotlight} (b)) arrives at the radar receiver simultaneously, thus resulting in the integral of the reflectivities on the line. By leveraging the arrival times of reflected radar signals, the radar receiver can obtain the integrated signal on the corresponding direction. According to the projection-slice theorem, the radar receiver can attain the slice of 2-dimensional Fourier transform $G$ of the reflectivities in the illuminated region, by fitting the spectrum $H$ of the integrated signal to the origin. If the aircraft changes multiple directions, it obtains the values of the 2-dimensional Fourier transform along multiple slicing lines passing the original. When sufficiently many slices have been obtained, the reflectivities of the illuminated region, namely the image, can be obtained from inverse Fourier transform.

\begin{figure}
  \centering
  \includegraphics[scale=0.4]{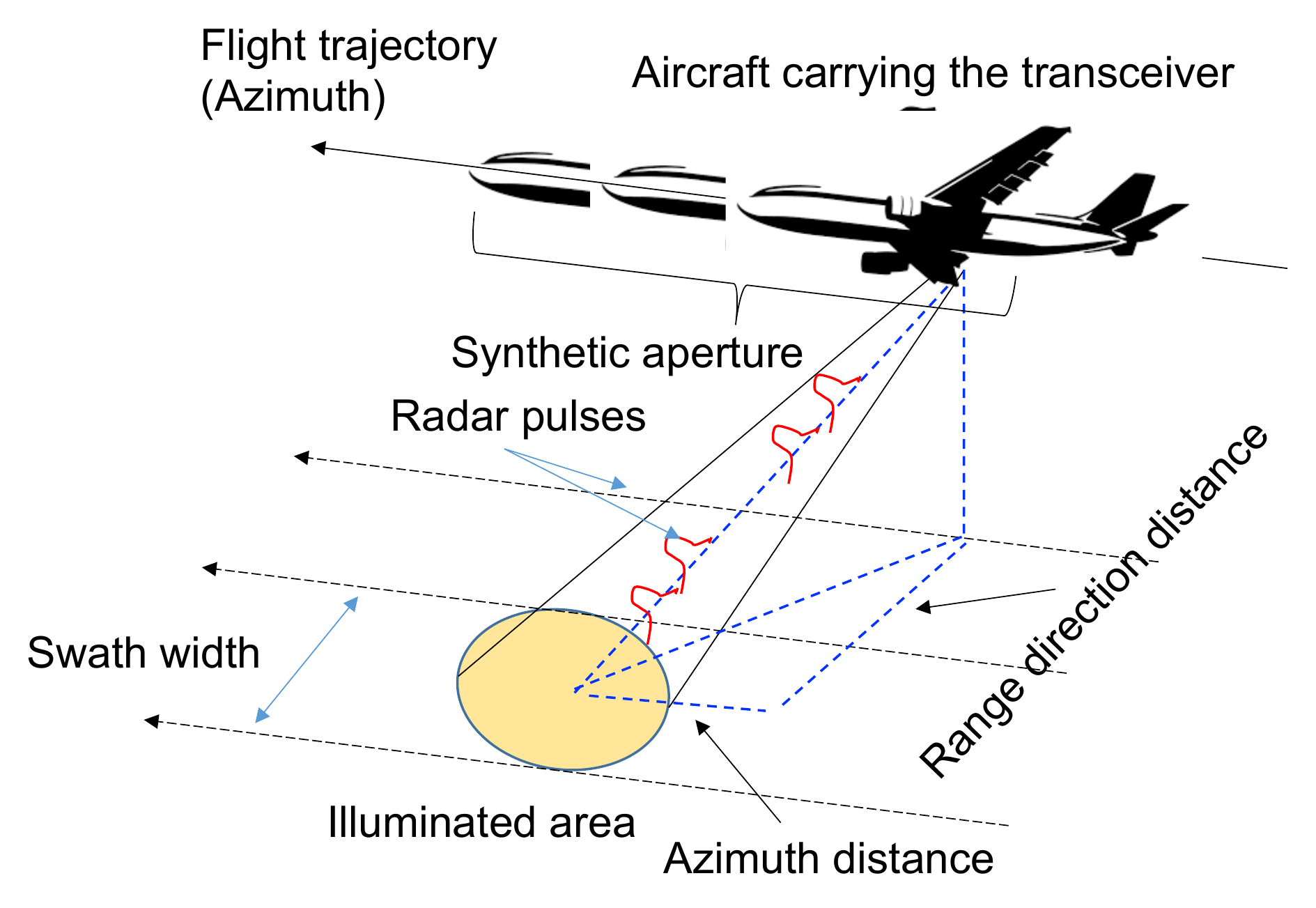}
  \caption{Working mechanism of scanning SAR.}\label{fig:scanning}
\end{figure}

Besides the spotlight SAR, scanning SAR can also be used for imaging, as illustrated in Fig. \ref{fig:scanning}. A moving object (aircraft or satellite) carries a transceiver, sheds radar pulses to a certain region, and receives the reflected pulses for imaging. The trick is that the same radar transceiver at different positions (thus different time) can be considered as multiple different antennas, thus forming a synthesized aperture (antenna) and improving the imaging resolution due to the enlarged aperture size. For locationing a significant reflector in the illuminated region, the SAR estimates the range direction distance by calculating the round trip time of radar pulses, and azimuth distance by evaluating the Doppler shift due to the relative movement to/from the reflector. In the proposed communication network based imaging, one can consider all the antennas of the receiving base stations as a synthetic aperture, whose geometric size becomes much greater than that of a single base station. On the other hand, different from the traditional SAR system, the base stations in the cellular network do not move, thus eliminating the possibility of the Doppler shift based azimuth distance estimation; moreover, the receiving base stations are randomly distributed, which is different from the smooth flight trajectory of the traditional SAR. This significant distinction requires a novel framework for the imaging using cellular networks. 

\subsection{Principle of ISAR}

\begin{figure}
  \centering
  \includegraphics[scale=0.5]{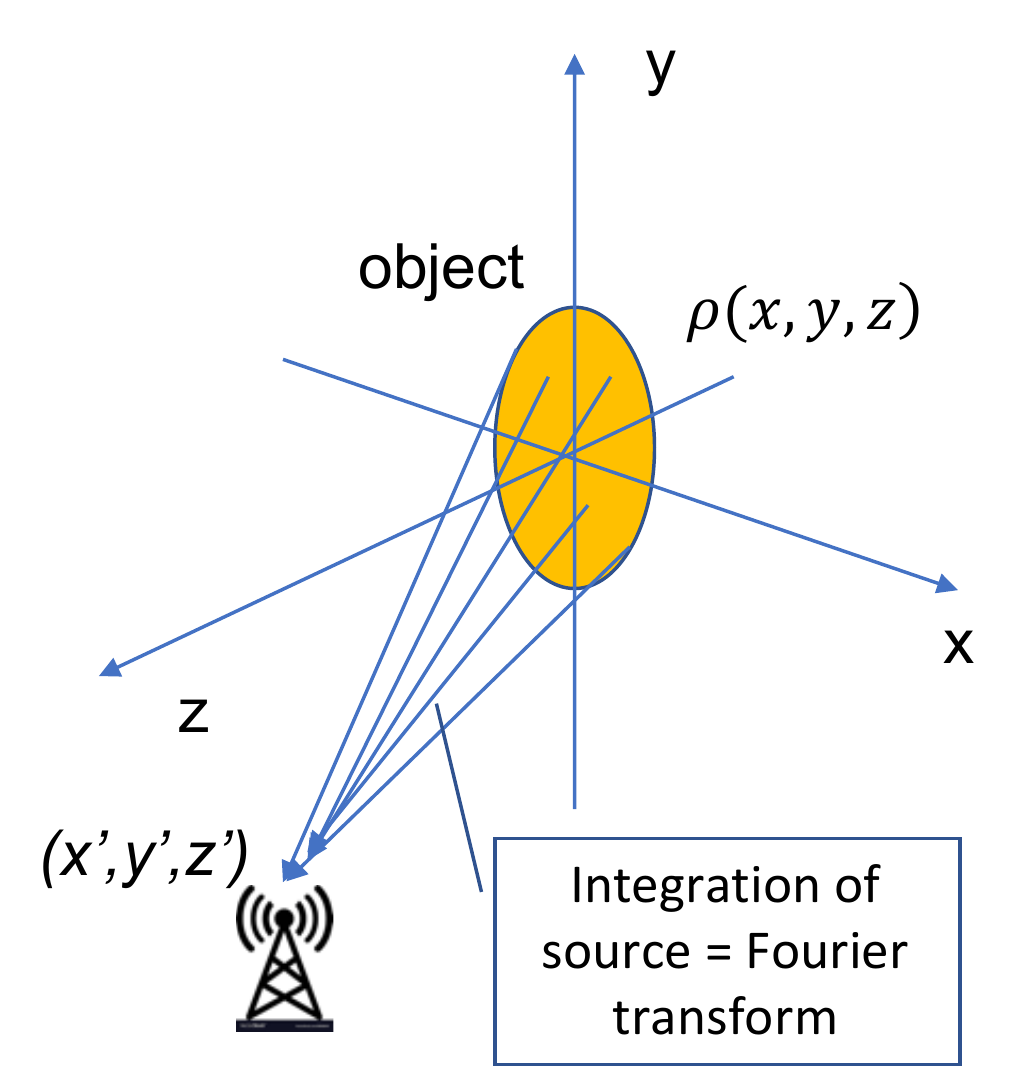}
  \caption{An illustration of ISAR.}\label{fig:ISAR}
\end{figure}

An alternative approach of imaging is to leverage the idea of inverse SAR (ISAR) imaging \cite{Sullivan2000}, in which the object moves while the antennas are fixed. The scenario is illustrated in Fig. \ref{fig:ISAR}. The relative movement of the antennas form multiple virtual antennas and thus the synthesis aperture. Again, denote the reflectivity of the illuminated region by $g(x,y,z)$, where $(x,y,z)\in \Omega$ and the origin is contained in $\Omega$. Then, it is shown that, in the far field, the received signal at position $(x',y',z')$ is given by
\begin{eqnarray}
x(x',y',z')=A\int_{\Omega}\rho(x,y,z)e^{-j(k_xx+k_yy+k_zz)}dxdydz,
\end{eqnarray}
where $k_x=kx'$ ($k$ is the wave number and equals the reciprocal of wavelength), $k_y=ky'$ and $k_z=kz'$. An interesting observation is that the received signal, as a function in the wave number domain, is the 3-dimensional Fourier transform of the source. In ISAR, the object is rotated, such that samples in the wavenumber space are obtained and thus the source $\rho$ is reconstructed by using the inverse Fourier transform. In the near field, the Weyl transformation can be used to modify the wave numbers \cite{Devaney2012}.

\section{Spotlight-SAR-Like Imaging}\label{sec:SAR}
In this section, we propose an imaging algorithm using multiple base stations and the principle of spotlight SAR. The main difference between the conventional spotlight SAR and the communication network based imaging is as follows:
\begin{itemize}
\item The aircraft for spotlight SAR moves following a trajectory, while the base stations are stationary.
\item The aircraft is both the transmitter and receiver, while in the communication network the transmitter and receivers are different base stations. 
\item In SAR systems, usually pulses or frequency modulation continuous waveforms (FMCW) are used. These waveforms are not used in communication systems; instead, spread spectrum signals such as orthogonal frequency division multiplexing (OFDM) are used. 
\end{itemize}
Therefore, the principle of spotlight SAR cannot be employed directly. As will be seen, the imaging procedure can be modified to fit the context of communication network based sensing, by following the fundamental mathematical principle of slicing and interpolation. 

\subsection{Slicing}
\begin{figure}
  \centering
  \includegraphics[scale=0.45]{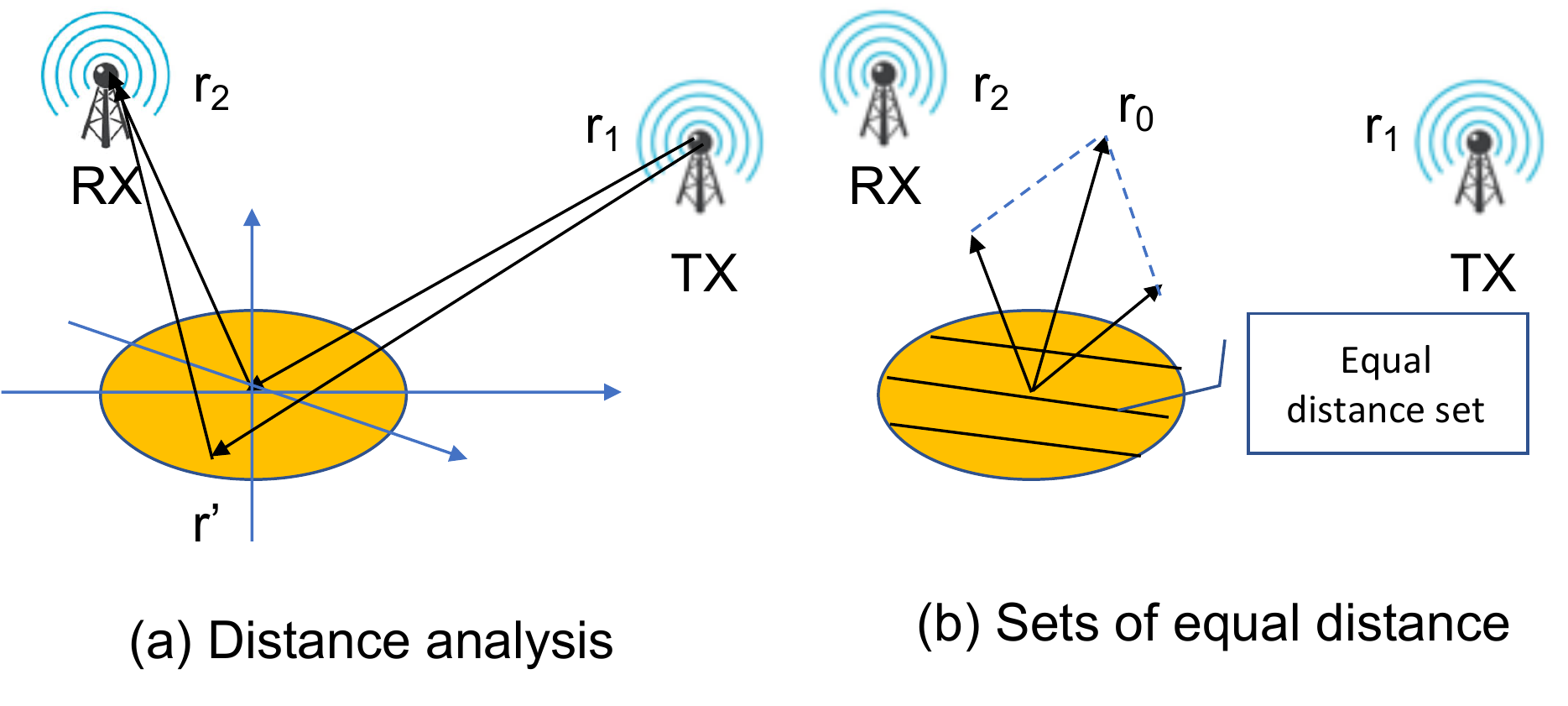}
  \caption{Calculation of distance.}\label{fig:distance}
\end{figure}

Following the principle of spotlight SAR, our first step is to estimate the spectrum slices of the illuminated area. For simplicity, we assume that the origin is located within the illuminated area. Consider a point $\mathbf{r}'$ in the illuminated area. Assume that the positions of the transmitter and receiver are $\mathbf{r}_1$ and $\mathbf{r}_2$, respectively. The distance from an illuminated point $\mathbf{r}'$ to the transmitter is given by
\begin{eqnarray}
\|\mathbf{r}'-\mathbf{r}_1\|\approx \|\mathbf{r}_1\|+\mathbf{r}'\cdot \frac{\mathbf{r}_1}{\|\mathbf{r}_1\|},
\end{eqnarray}
where the approximation is valid when $\|\mathbf{r}_1\|\gg \|\mathbf{r}'\|$. Similarly we have
\begin{eqnarray}
\|\mathbf{r}'-\mathbf{r}_2\|\approx \|\mathbf{r}_2\|+\mathbf{r}'\cdot \frac{\mathbf{r}_2}{\|\mathbf{r}_2\|}.
\end{eqnarray}

Therefore the traveling distance of the signal sent from the transmitter, reflected at $\mathbf{r}'$, and received by the receiver, is approximated by
\begin{eqnarray}
&&\|\mathbf{r}'-\mathbf{r}_1\|+\|\mathbf{r}'-\mathbf{r}_2\|\nonumber\\
&\approx& \|\mathbf{r}_1\|+\|\mathbf{r}_2\|+\mathbf{r}'\cdot \left(\frac{\mathbf{r}_1}{\|\mathbf{r}_1\|}+\frac{\mathbf{r}_2}{\|\mathbf{r}_2\|}\right).
\end{eqnarray}
Hence, the set of reflection points that result in the same traveling time is given by
\begin{eqnarray}
\mathbf{r}'\cdot \left(\frac{\mathbf{r}_1}{\|\mathbf{r}_1\|}+\frac{\mathbf{r}_2}{\|\mathbf{r}_2\|}\right)=const,
\end{eqnarray}
which is an approximately the straight line perpendicular to the direction 
\begin{eqnarray}\label{eq:r0}
\mathbf{r}_0=\frac{\mathbf{r}_1}{\|\mathbf{r}_1\|}+\frac{\mathbf{r}_2}{\|\mathbf{r}_2\|}, 
\end{eqnarray}
as illustrated in Fig. \ref{fig:distance} (b). Hence, a measurement at the receiver collects the integrated signals from the equal distance lines. 

\subsection{Impulsive Response}
When the signal sent from the transmitter is a delta function $\delta(t)$, the received signal is exactly the accumulated signal $h(t)$ along the direction specified by $\mathbf{r}_0$, namely
\begin{eqnarray}
h(t)\propto \int_{\|\mathbf{r}_1\|+\|\mathbf{r}_2\|+\mathbf{r}'\cdot \mathbf{r}_0=ct} g(\mathbf{r}')d\mathbf{r}',
\end{eqnarray}
which is the accumulated reflectivity along the equal distance line (recall that $g$ is the reflectivity of the illuminated point). The profile of $h(t)$ can be used to reconstruct the slice of Fourier transform of the charge density of the illuminated region, by leveraging the projection-slice theorem. However, different from SAR systems, the communication systems do not use pulse waveform or FMCW, which can obtain $h(t)$ directly (e.g., sampling the reflected pulses). Therefore, we consider the subsequent two approaches that are essentially equivalent but take different paths to the desired result.

\subsubsection{Spatial Fourier Transform} 
\begin{figure}
  \centering
  \includegraphics[scale=0.55]{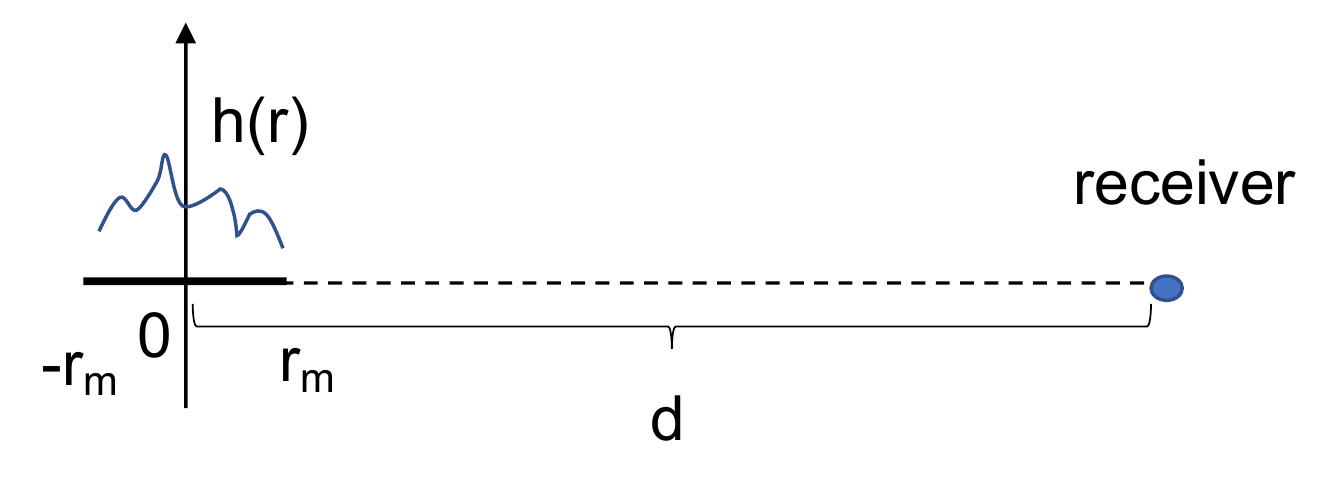}
  \caption{An illustration of the virtual source $h(r)$.}\label{fig:one_dim}
\end{figure}

We first notice that the impulse response $h(t)$ is indeed also a function of space, which can be written as $h(r)$ (without changing the notation for simplicity), where $r=ct$ is the EM wave propagation distance. Therefore, we can consider $h(r)$ as a virtual source charge distribution, supported within $(-r_m,r_m)$, where $2\delta r$ is the range of the propagation distance and the distance between the source center and the receiver is $d$, as illustrated in Fig. \ref{fig:one_dim}. Given a single-tone carrier input with wavenumber $k=\frac{1}{\lambda}$ and without modulation, the received signal is given by
\begin{eqnarray}
H(t,k)&=&\int_{-r_m}^{r_m} h(r)e^{-j((r+d)k+wt)}dr\nonumber\\
&=&e^{-jwt}e^{-jdk}\int_{-r_m}^{r_m} h(r)e^{-jrk}dr.
\end{eqnarray}
By demodulation (to remove $e^{-jwt}$) and phase alignment $e^{-jdk}$, we obtain the complex scalar:
\begin{eqnarray}
H(k)=\int_{-r_m}^{r_m} h(r)e^{-jrk}dr,
\end{eqnarray}
which shows that the received signal $H(k)$ is the spatial Fourier transform of $h(r)$. Notice that $H(k)$ is simply the received signal at the corresponding subcarrier with frequency $ck$. Therefore, we use the received signals $\{H_m\}_{m=1,...,M}$ on the different subcarriers as the samples of $H(k)$ in the wavenumber space, as the samples in the spatial spectrum of the reflectivity of the illuminated area. Then, the samples $\{H_m\}_{m=1,...,M}$ can be applied to the projection-slice theorem. 

\begin{figure}
  \centering
  \includegraphics[scale=0.55]{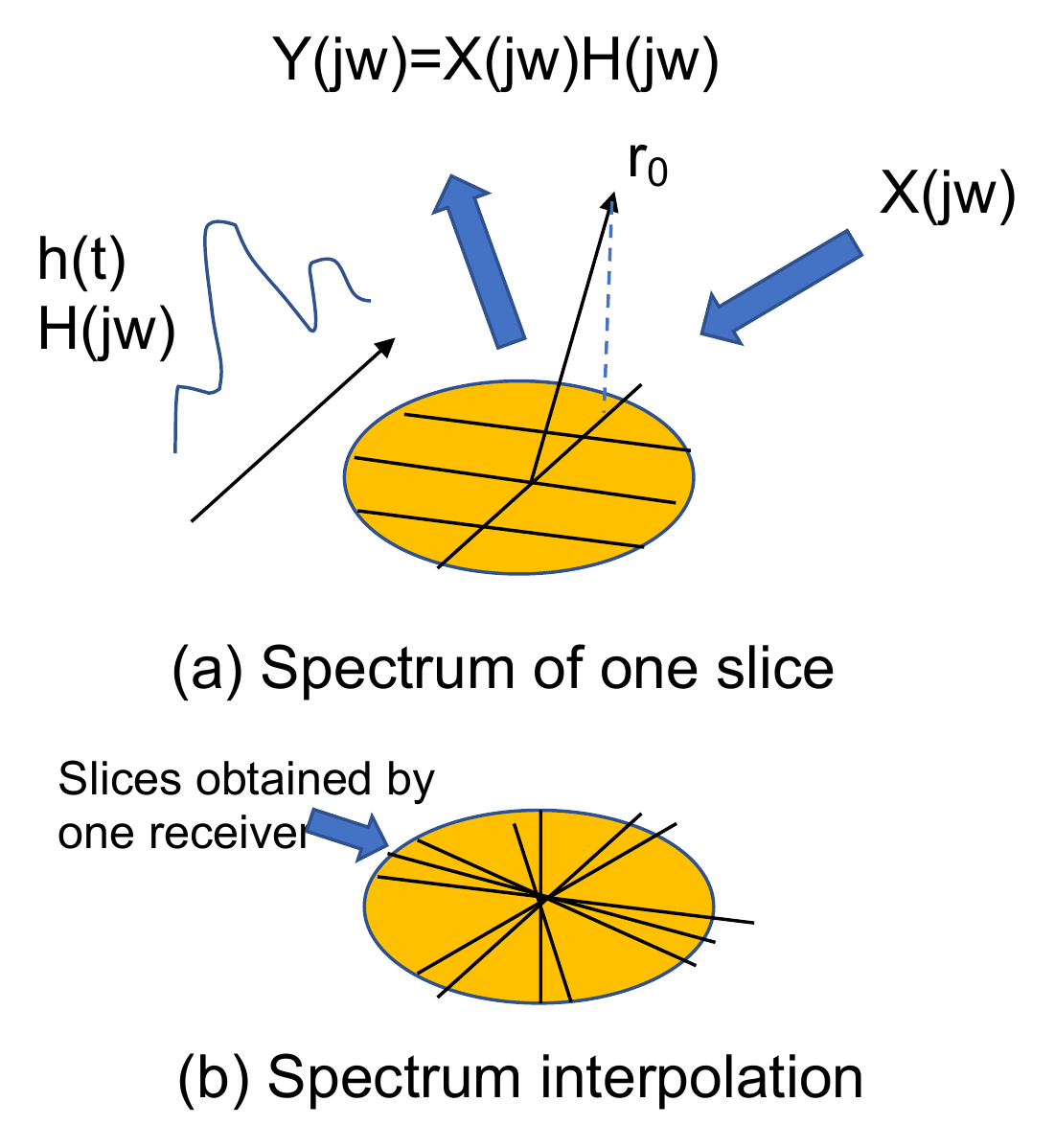}
  \caption{Slicing and interpolation in the spectrum.}\label{fig:slicing}
\end{figure}

\subsubsection{Frequency Response}
We can also consider the illumination procedure as a linear system, where $h(t)$ is the impulse response, while the input and output are the transmitted and received signals. The transfer function $H(jw)$ equals the Fourier transform of $h(t)$.
In practice, the transmitter signal cannot be an impulse $\delta(t)$. When the spectra of the transmitted and received signals are $X(jw)$ and $Y(jw)$, respectively, the transfer function $H$ can be estimated as $H(jw)=\frac{Y(jw)}{X(jw)}$. When OFDM is employed, we obtain the sample estimations of the transfer function:
\begin{eqnarray}\label{eq:freq_response}
H(jw_m)=\frac{Y_m}{X_m}, \qquad m=1,...,M,
\end{eqnarray}
where $w_m$ is the frequency of the $m$-th subcarrier, $Y_m$ and $X_m$ are the received and transmitted signals on the $n$-th subcarrier. This is illustrated in Fig. \ref{fig:slicing} (a). Notice that $H(jw_m)$ is simply $H_m$ in the above discussion. Therefore, the two approaches are essentially the same.

\subsubsection{Data Collection Surface}

\begin{figure}
  \centering
  \includegraphics[scale=0.55]{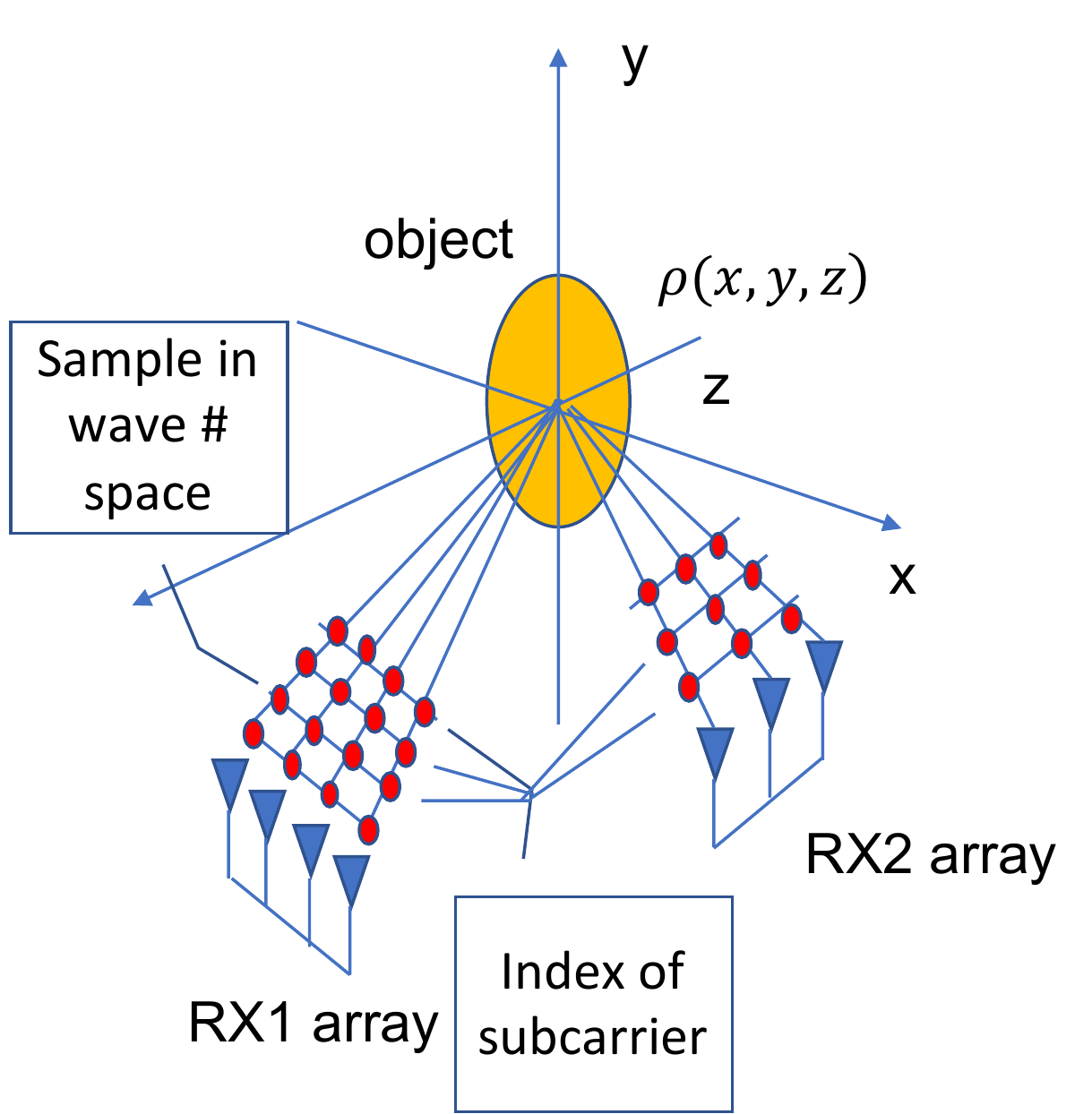}
  \caption{Sampling in the wave number space.}\label{fig:IDFT}
\end{figure}

Since we assume multiple receive antennas at each receiver, each antenna represents a slice of the Fourier transform domain of the reflectivity in the illuminated area. When the antenna arrays are linear, the data can be intuitively represented by data collection surfaces, illustrated in Fig. \ref{fig:IDFT}, where the two data collection surfaces of two different receiving base stations are shown. The cross side of the surface is the direction of the antennas, while the range side means the wave number dimension. We denote by $H^n_{ml}$ the spatial-spectrum data on the $l$-th subcarrier at the $m$-th antenna of the $n$-th base station. The data form a grid on the data collection surface. We denote by $X$ the cross axis and $Y$ the ranging axis. Therefore, the data can be denoted by $H^n(X_l,Y_m)$. 

\begin{figure}
  \centering
  \includegraphics[scale=0.55]{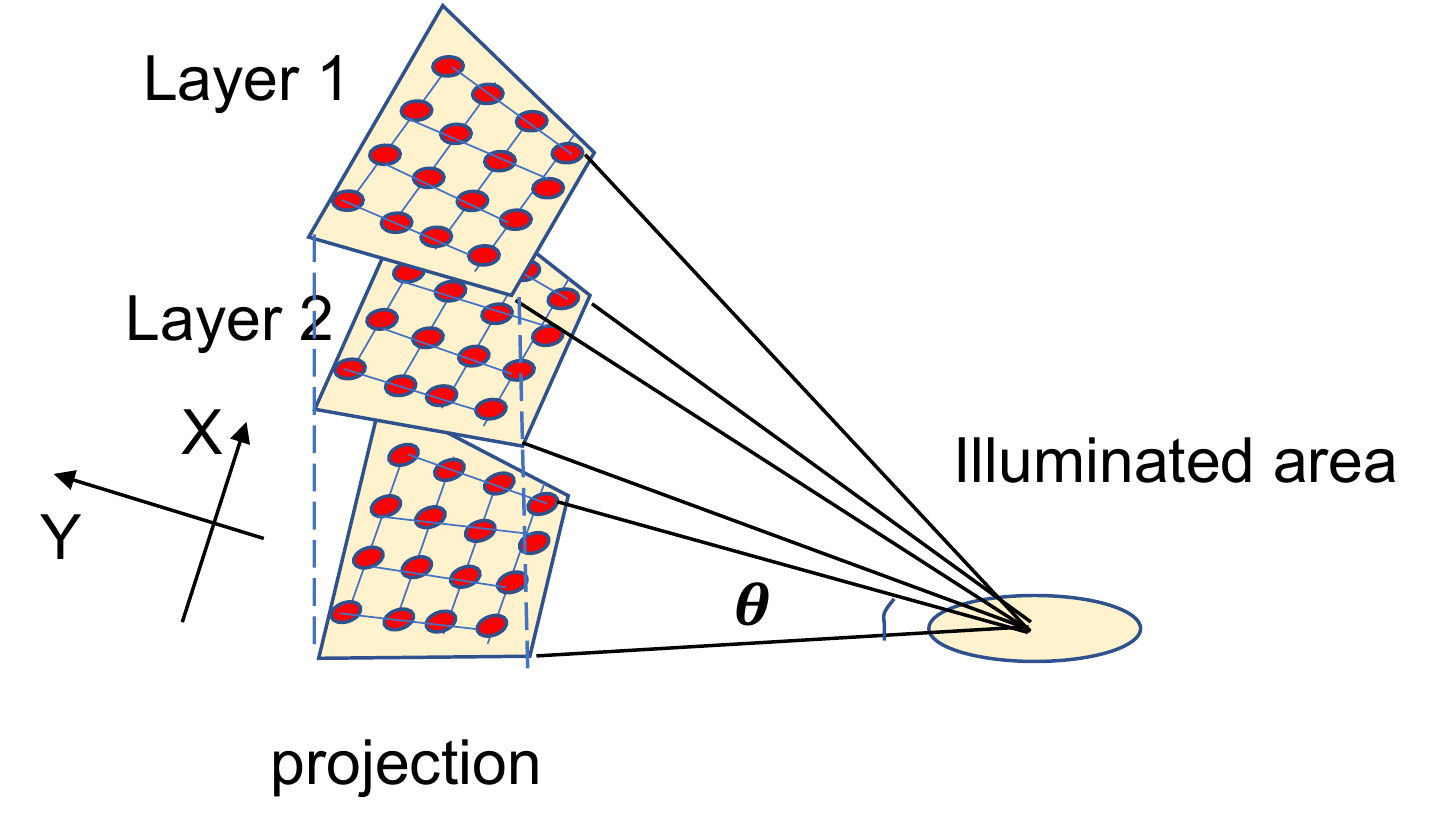}
  \caption{Projection of multiple layers into one data collection surface.}\label{fig:projection}
\end{figure}

When the antenna arrays are two-dimensional with multiple horizontal layers, the data collection surface becomes a data collection volume. We can project the multiple surfaces into a ground one, as illustrated in Fig. \ref{fig:projection}, where the antenna has two layers with different heights. After the projection, the equivalent data in the ground data collection surface is given by
\begin{eqnarray}
\mathbf{H}=\{H^{np}(X_{l},Y_{m}\cos\theta_p)\}_{p,l,m},
\end{eqnarray}
where $p$ is the index of the layer and $\theta_p$ is the angle between the $p$-th surface and the ground. In subsequent discussion, we assume that each receiving base station has projected the data into one ground data collection surface. 

\subsection{Missing Data}
If the data $H^n_{ml}$ is known for all wavenumbers, namely the data is supported along all the $Y$ axis and the data support on the $X$ axis is sufficiently long, the reflectivity image can be obtained by simply carrying out inverse Fourier transform. However, we only have limited data, and the data beyond the communication frequency band and the antenna size is missing, which will be assumed zero during the image reconstruction. In the subsequent discussion, we address the missing data. Note that, despite the missing data due to limited ranges in the $X$ and $Y$ axises, it has been demonstrated in practical SAR systems that the illuminated area can be well reconstructed \cite{Munson1984}. In our later numerical simulations, the same conclusion will be achieved.

\subsubsection{Frequency Offset in Range}
The base stations sample on only a bounded portion of the wavenumbers along the $Y$ axis. This is called the frequency offset issue in SAR, namely there is an offset between the origin of the spatial spectrum and the center of the data collection surface. 
To handle the frequency offset, we consider a single data collection surface. We shift the origin to the center of the data collection plane, which is denoted by $(0,Y_0)$. The shifted version of the spatial spectrum of the reflectivity is given by
\begin{eqnarray}
G_0(X,Y)&=&G(X,Y-Y_0)\nonumber\\
&=&\int g'(x,y)e^{-j(xX+yY)}dxdy,
\end{eqnarray}
where $g'(x,y)=g(x,y)e^{-jyY_0}$. Therefore, by shifting the center, we can recover $g'(x,y)$ which has different phases from the true values of $g(x,y)$. Notice that $|g'(x,y)|=|g(x,y)|$. Therefore, we can calculate $g'(x,y)$ since only the magnitude information is needed for imaging. Note that the $Y$ axis resolution is given by
\begin{eqnarray}
\rho_Y=\frac{c}{2W},
\end{eqnarray}
where $W$ is the bandwidth. Therefore, for a communication signal with a bandwidth of 300MHz, the resolution is 1 meter, which is reasonable. 

\subsubsection{Narrow Cross Bandwidth}
We also have concerns in the $X$ axis. In airborn SAR systems, the aircraft passes a long distance along the $X$ axis and thus yields a sufficient large data collection surface. However, in communication networks, the size of the base station antenna is very limited. The resolution along the $X$ axis is given by (Section 2.4, \cite{Jakowatz1999})
\begin{eqnarray}
\rho_X=\frac{\lambda}{2\delta\theta},
\end{eqnarray}
where $\delta\theta$ is the angle of the arc formed by the surface (as illustrated in Fig. \ref{fig:projection}). Take the 5GHz band for instance. If the $X$-width of the antenna array is 0.5m and the distance between the center of illuminated area and the antenna center is 100m, the resolution along the $X$ axis is 12m, which is prohibitive in practice. Therefore, the data collection surface resulted by a single base station is insufficient for the cross direction. To handle this challenge, we need to fuse the data from multiple base stations, which will be discussed subsequently.

\subsection{Alignment}
Different from the traditional SAR system, in which the received signals have very close spatial positions, the distributed imaging in the paper has distinct data patches in the spatial spectrum, which are collected from base stations of significantly different distances and positions. The position difference of the receivers results in different phase offsets and magnitude attenuations. Therefore, before the image reconstruction, the signals received at different positions will be aligned:
\begin{eqnarray}\label{eq:alignment}
\tilde{H}^n(X_l,Y_m)=H^n(X_l,Y_m)L(d_n)e^{-jd_nk_l},
\end{eqnarray}
where $d_n$ is the distance between the receiver and the illuminated area, which can be measured using the time elapse of signal, and $L$ is the path loss due to the distance $d_n$.

\begin{figure}
  \centering
  \includegraphics[scale=0.45]{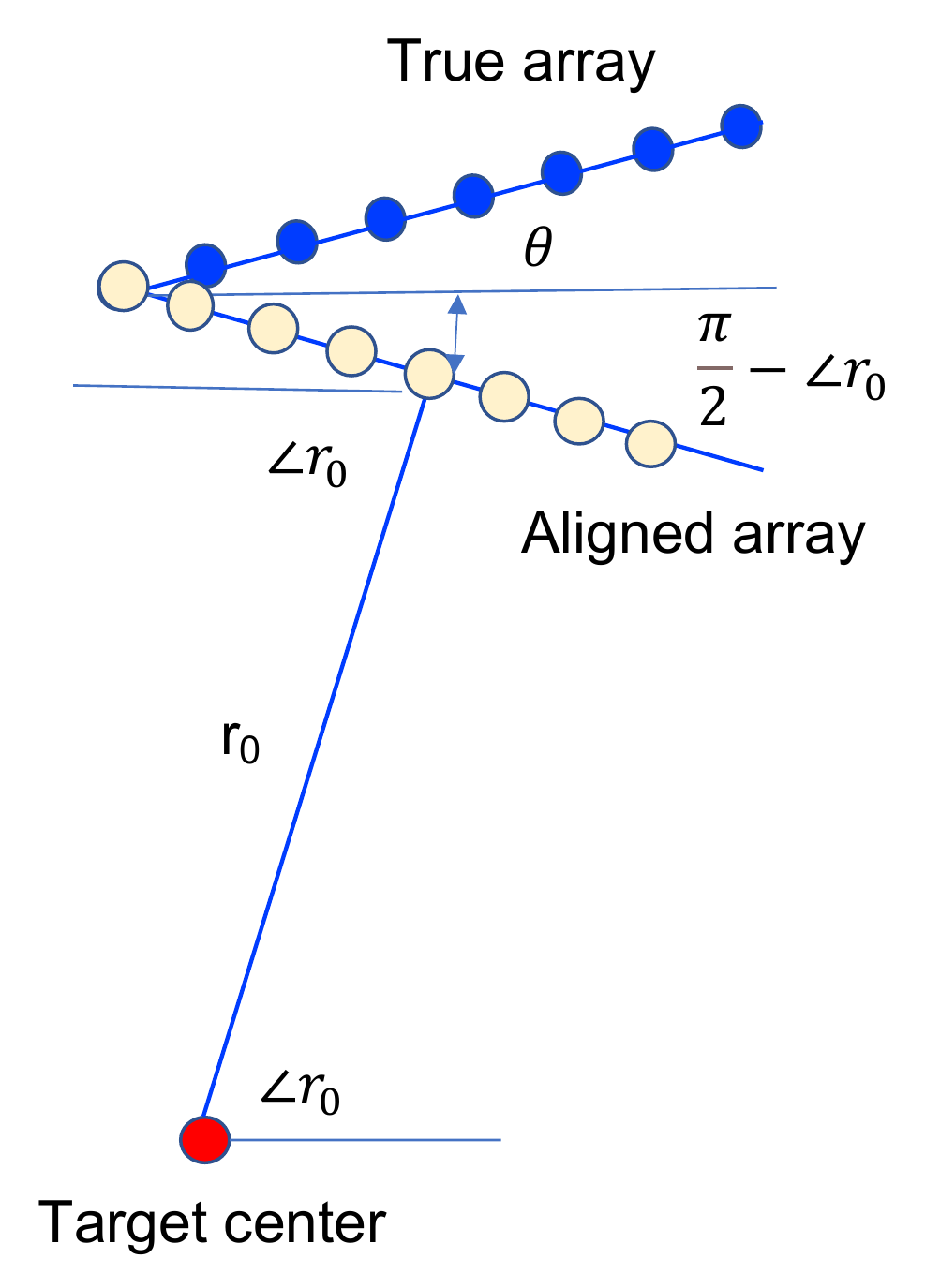}
  \caption{Rotation of antenna array for alignment.}\label{fig:rotate}
\end{figure}

Another thing to be aligned is the direction of the receive antenna array. In traditional SAR systems, the trajectory of the aircraft carrying the antenna (the $X$-axis) is almost perpendicular to the radius direction (the $Y$-axis), which is very natural to the SAR system. However, in the context of the network sensing, the direction of the antenna array is not necessarily perpendicular to the direction $\mathbf{r}_0$, which causes a linear distortion to the received signal phase. Therefore, the phase of the received signal needs to be aligned:
\begin{eqnarray}\label{eq:alignment2}
\tilde{H}^n(X_l,Y_m)=H^n(X_l,Y_m){e^{j (l-1)\delta d k_m \sin(\psi)}},
\end{eqnarray}
where the angle $\psi$ is given by
\begin{eqnarray}\label{eq:angle}
\psi=\frac{\pi}{2}-\angle \mathbf{r}_0+\theta,
\end{eqnarray}
which can be easily seen in Fig. \ref{fig:rotate}.

\subsection{Reconstruction by Irregular Inversion}

\begin{figure}
  \centering
  \includegraphics[scale=0.45]{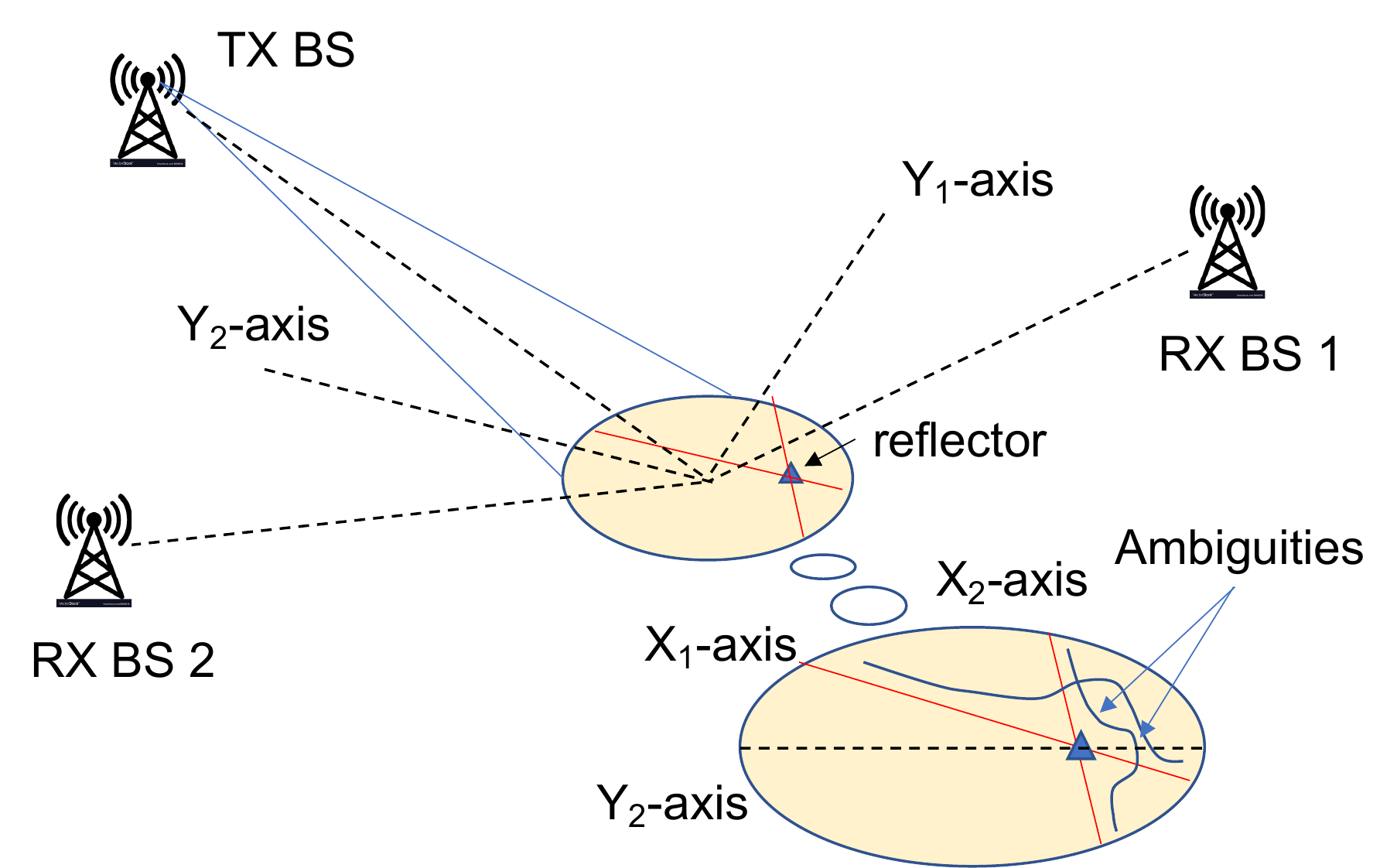}
  \caption{The fusion of data at different receiving base stations.}\label{fig:fusion}
\end{figure}

As discussed above, in the context of communication network based imaging, the cross bandwidth of a single receiving base station is too narrow, which results in substantial ambiguity in the cross direction ($X$-axis). Therefore, multiple receiving base stations are needed for reasonably good imaging. The ambiguity of one base station can be resolved by the measurements of another base station. This is illustrated in Fig. \ref{fig:fusion}. Consider a significant reflector in the illuminated region. From the measurement at base station 1, the location of the reflector is ambiguous, since its signal is mixed with the reflections along the $X_1$ axis and cannot be distinguished due to the poor cross resolution at base station 1. When base station 2 also receives the reflected signal, it has a good resolution along the $Y_2$ axis, which helps to locate the reflector and significantly resolves the reflector. 

\begin{figure}
  \centering
  \includegraphics[scale=0.45]{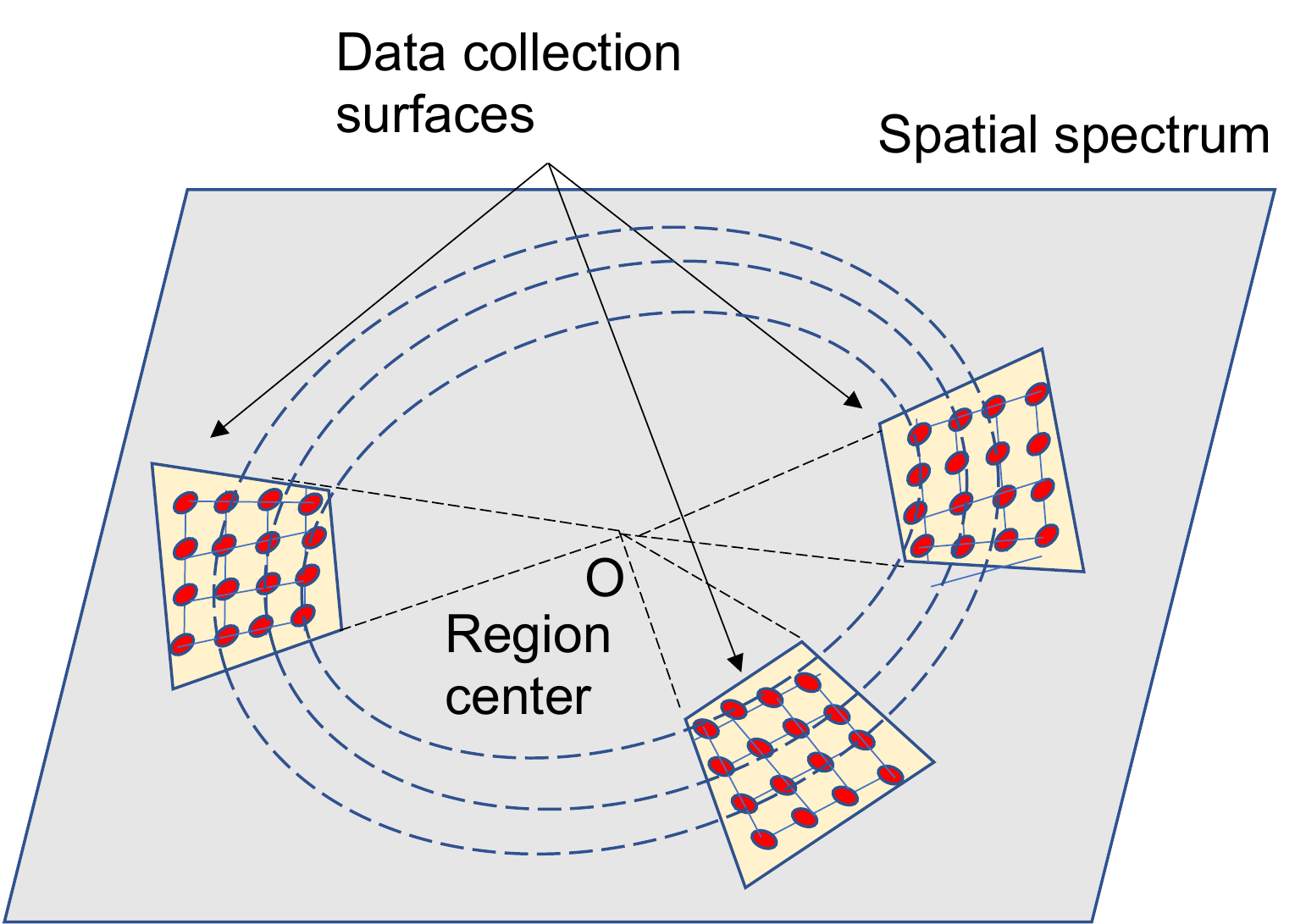}
  \caption{The fusion of data patches.}\label{fig:patches}
\end{figure}

Based on the above intuition, we fuse the measurements at different base stations for imaging. As illustrated in Fig. \ref{fig:patches}, we collect the data in the spatial spectrum at different base stations. Therefore, the role of the receiving base stations is to illuminate and probe multiple `patches' in the spatial spectrum. The spatial spectrum out of these patches are either assumed to be zero or interpolated. 

The detailed fusion algorithm is given in Procedure \ref{alg:fuse}.
\begin{algorithm}[H]
	\caption{}\label{alg:fuse}
	\begin{algorithmic}[1]
	        \STATE{Set an $2S\times 2S$ all-zero matrix $\tilde{H}$.}
		\FOR{Each receiving BS}
		        \FOR{Each antenna of the BS}
   		                \STATE{Calculate the direction $\mathbf{r}_0=(r_{01},r_{02})$ using (\ref{eq:r0}).}
		                \STATE{Calculate the angle $\theta=\tan^{-1}(r_{02}/r_{01})$.}
		                \STATE{Project the measurement vector $H(1:M)$ on the subcarriers to the plane, obtain $H'$.}
		                \FOR{Each measurement $H'_k$ in $H'$}
		                         \STATE{Calculate the index $\mathbf{z}=([k\cos\theta],[k\sin\theta])$.}
		                         \STATE{Set $H'(\mathbf{z})=H'_k.$}
		                         \STATE{Align the signals at different base stations using (\ref{eq:alignment}).}
		                \ENDFOR
		        \ENDFOR
		\ENDFOR
		\STATE{Carry out inversion $\tilde{H}$ to obtain the image reconstruction.}
	\end{algorithmic}
\end{algorithm}

\subsection{Reconstruction by IDFT and Fusing}
In the above inversion algorithm, a large matrix needs to be inversed, which results in substantial computational cost. In traditional SAR systems, the spatial samples in the image can be easily computed using IDFT, since the measurements collected in the data patches are the Fourier transform of the image. Although the data patch is only a portion of the spatial spectrum, the spectrum center can be shifted to the center of the patch for IDFT. Different from the traditional SAR system in which there is only one data patch, we are facing multiple isolated data patches in the communication network based imaging. One approach is to place the center of spatial spectrum at the origin and carry out the IDFT for the data patches by nullifying the samples without measurements. However, the dimension of the spatial spectrum will be substantially increased, since the frequencies not in the signal band need to be incorporated. For example, consider a 128-subcarrier system with frequency spacing 128kHz, carrier frequency 5GHz and 64 antennas, the IDFT will be a $128\times 64$ one if the center is shifted to the data patch, while it becomes $39190\times 64$ if keeping the origin. Therefore, it is prohibitively difficult to pool all the data patches and carry out a single IDFT.

Therefore, in this paper, we propose IDFT-based inversion and fusion for the distinct patches. For each data patch, we rotate the orientation to the direction $\mathbf{r}_0$, and shift the center of the spatial spectrum to that of the data patch. The spatial spectrum is normalized by the offset of the minimum of $\mathbf{k}\cdot\mathbf{r}$, which is for the application of DFT:
\begin{eqnarray}
\bar{H}^n_{lm}=\bar{H}^n_{lm}e^{j \min_{\mathbf{r}}(\mathbf{k}\cdot\mathbf{r})}.
\end{eqnarray}

After the shift of center, the spatial spectrum at base station $n$ sampled at the grid $\{m\delta k_1, n\delta k_2\}_{m=1,...,M,n=1,...,N_a}$, is given by
\begin{eqnarray}
\bar{H}^n_{lm}&\propto& \int g(\mathbf{r})e^{-i\mathbf{k}_{lm}\cdot\mathbf{r}}d\mathbf{r}\nonumber\\
&\approx &\sum_{a,b}g_{a,b}e^{-i(al\delta k_1\delta r_1+bm \delta k_2\delta r_2)}\nonumber\\
&=&\sum_{a,b}g_{a,b}e^{-i\left(\frac{al}{M}+\frac{bm}{N_a}\right)},
\end{eqnarray}
where the step parameters $\delta k_1$, $\delta k_2$, $\delta r_1$ and $\delta k_2$ are chosen such that $M=\frac{1}{\delta k_1\delta r_1}$ and $N_a=\frac{1}{\delta k_2\delta r_2}$, in order for the 2-dimensional inverse DFT. 

\begin{figure}
  \centering
  \includegraphics[scale=0.35]{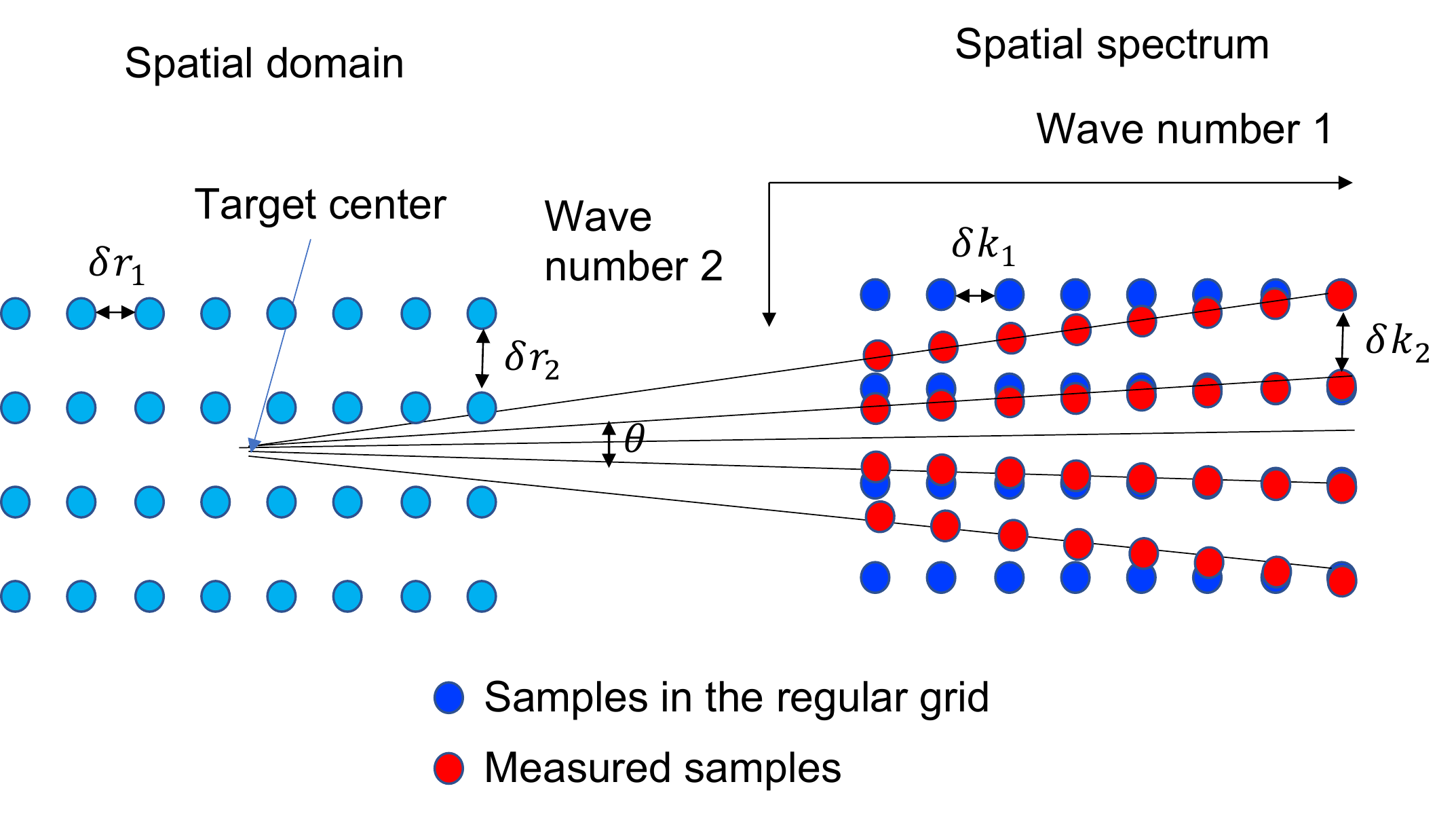}
  \caption{The setup of 2-dimensional DFT and IDFT.}\label{fig:inverse}
\end{figure}

We select the step parameters $\delta k_1$, $\delta k_2$, $\delta r_1$ and $\delta k_2$ as follows:
\begin{itemize}
\item We select $\delta k_1$ and $\delta k_2$ such that the samples at $\{m\delta k_1, n\delta k_2\}_{m=1,...,M,n=1,...,N_a}$ are close to the measured samples $\tilde{H}_{mn}$:
\begin{eqnarray}\label{eq:deltak}
\left\{
\begin{array}{ll}
\delta k_1=\frac{\delta f}{c}\cos\left(\frac{\theta}{2}\right)\\
\delta k_2=\frac{\delta f}{c}\sin\left(\frac{\theta}{2}\right)
\end{array}
\right..
\end{eqnarray}
\item We select 
\begin{eqnarray}\label{eq:deltar}
\left\{
\begin{array}{ll}
\delta r_1=\frac{1}{M\delta k_1}\\
\delta r_2=\frac{1}{N_a\delta k_2}
\end{array}
\right..
\end{eqnarray}
\end{itemize}

The spatial spectrum samples $\{\bar{H}^n_{lm}\}_{lm}$ are obtained from the interpolations of the measured samples $\{\tilde{H}^n_{lm}\}_{lm}$. Then, the spatial samples of the images at positions $\{a\delta r_1,b\delta r_2\}$ are obtained from the inverse DFT of the spatial spectrum samples at $\{\bar{H}^n_{lm}\}_{lm}$.  

Once the reconstructed images, of the same illuminated area, have been obtained from distinct data patches, we can fuse them into the same image by interpolating among the reconstructed samples. The procedure is summarized in Procedure \ref{alg:fuse2}.

\begin{algorithm}[H]
	\caption{}\label{alg:fuse2}
	\begin{algorithmic}[1]
		\FOR{Each receiving BS}
   	                \STATE{Calibrate the received signals.}
	                \STATE{Rotate the orientation and shift the center.}
	                \STATE{Determine the parameters $\delta k_1$, $\delta k_2$, $\delta r_1$ and $\delta r_2$ using (\ref{eq:deltak}) and (\ref{eq:deltar}).}
                         \STATE{Carry out interpolation for the spatial spectrum samples at regular grid.}
                         \STATE{Carry out IDFT to obtain the spatial image.}		
		\ENDFOR
		\STATE{Carry out interpolation using the reconstructed images.}
	\end{algorithmic}
\end{algorithm}

\subsection{Reconstruction with Insufficient Cross Resolution}\label{sec:insuff}
A major challenge to the communication network based imaging is the cross direction resolution. In traditional SAR, the cross resolution is assured by the long flight trajectory of aircraft. However, in the proposed communication network based imaging, for each base station, the cross direction span is determined by the size of antenna array. When the antenna spacing is $\frac{\lambda}{2}$, the cross direction resolution is 
\begin{eqnarray}
\delta x\approx \frac{\lambda}{\frac{2N_a\lambda}{2d}}=\frac{d}{N_a},
\end{eqnarray}
which equals 1.56m when $d=100m$ and $N_a=64$. The resolution will be even worse when $N_a$ is small. In such a scenario, only the range direction information (range and width) is obtained from the Fourier transform. When there are two base stations receiving the signal, we can use the intersection of the perpendicular lines to locate the significant reflector.

\begin{figure}
  \centering
  \includegraphics[scale=0.45]{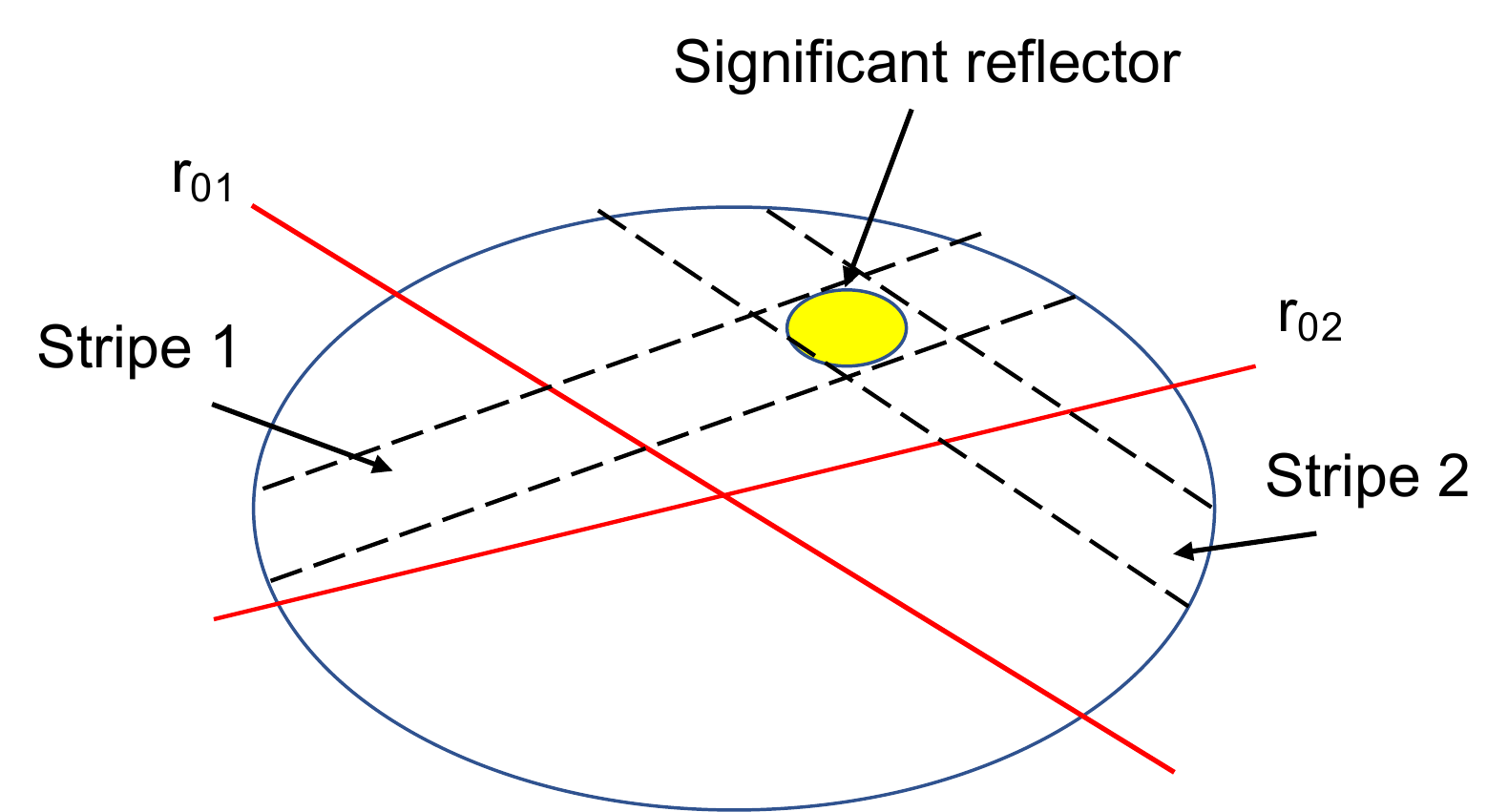}
  \caption{Estimation of significant reflector with the lack of cross resolution.}\label{fig:crossit}
\end{figure}

\subsection{3-dimensional Reconstruction}
The above imaging procedure considers only two-dimensional image and ignores the vertical information. Based on the 2-dimensional imaging, we propose a 3-dimensional imaging procedure, based on the assumption that only reflection on the surface exists and there is no EM wave penetration. Note that the different heights of different base stations, and the possible different layers of antennas in the same base station can provide the resolution in the vertical direction.  

We assume that the surface height is given by $z=h(x,y)$. Then, the 3-dimensional Fourier transform of the reflectivity is given by
\begin{eqnarray}
G(X,Y,Z)=\int r(x,y)e^{-jZh(x,y)}e^{-j(xX+yY)}dxdy.
\end{eqnarray}
We select a series of $\{Z_i\}$ and project the data collection surfaces to the heights $\{Z_i\}$.
By using the above the 2-dimensional imaging, we obtain $r_i(x,y)=r(x,y)e^{-jZ_ih(x,y)}$.
For the ground plane $Z=0$, we obtain $r(x,y)$. Thus for each $Z_i$ we obtain
\begin{eqnarray}
e^{-jZ_ih(x,y)}=\frac{r_i(x,y)}{r(x,y)}.
\end{eqnarray}
Hence, for each $(x,y)$, we obtain the data $\{e^{-jZ_ih(x,y)}\}_i$. Here $h(x,y)$ can be considered as the frequency of the complex exponential function $f(Z)=\left\{e^{-jZh(x,y)}\right\}_i$. Due to the missing data, the collection of $\left\{e^{-jZh(x,y)}\right\}_i$ will be non-ideal. Therefore, we carry out DFT for the data set $\left\{e^{-jZh(x,y)}\right\}_i$ and pick the spectrum peak frequency as $h(x,y)$, thus obtaining an estimation of the surface height.

\section{ISAR Imaging}\label{sec:ISAR}
In this section, we use the principle of ISAR imaging for the reconstruction of the illuminated area in the communication networks. Note that both the ISAR and spotlight SAR imaging approaches are essentially based on sampling in the spatial spectrum after Fourier transform. While the spotlight SAR provides a more intuitive explanation (e.g., the resolutions in the range and cross directions), the ISAR approach provides a framework for 3-dimensional image reconstruction.

When using the ISAR principle of imaging, the signals received at different antennas and different subcarriers can be considered as samples in the wavenumber space. For example, when an antenna is located at $(x',y',z')$, the received signal at a subcarrier with frequency $f$ is a sample at $(\frac{2\pi fx'}{c},\frac{2\pi fy'}{c},\frac{2\pi fz'}{c})$. For a 1-dimensional linear antenna array with $N$ antennas and $M$ subcarriers, the received signals are $MN$ samples on one plane, which is determined by the antenna array direction and the center of illuminated region, which is illustrated in Fig. \ref{fig:IDFT}. 

Since we cannot obtain all the values in the wavenumber space, we can only use the samples to infer the three dimensional spectrum of the charge densities. Due to the availability of the discrete samples, we also discretize the space of the target. Suppose that we partition the target to an $M\times M\times M$ grid with a spacing $\delta$. Therefore, we define
\begin{eqnarray}
\rho^{l,m,n}=\rho(l\delta,m\delta,n\delta),
\end{eqnarray}
for $(l,m,n)\in \{-M/2,...,M/2\}^3$. Suppose there are $K$ antennas, whose positions are given by the set of coordinates $\{\mathbf{r}_k\}_{k=1,..,K}$. Then, the field $x$ and the source $\rho$ are related by 
\begin{eqnarray}
x^s=\rho^{l,m,n}\Lambda^s_{l,m,n},
\end{eqnarray}
where we use the Einstein summation notation and the $(1,3)$ tensor $\Lambda^s_{l,m,n}$ is given by
\begin{eqnarray}
\Lambda^s_{l,m,n}=A e^{-j(kx_sl\delta+ky_sm\delta+kz_sn\delta}.
\end{eqnarray}

To recover $\rho$, we calculate the inverse of $\Lambda^s_{l,m,n}$, denoted by $\Gamma^{l,m,n}_s$ and obtain
\begin{eqnarray}
\rho^{l,m,n}=\Gamma_s^{l,m,n}x^s.
\end{eqnarray}

\section{Performance Evaluation}\label{sec:performance}
In this section, we analyze the performance of the image reconstruction in the communication network. For simplicity, we focus on only the spotlight SAR approach.

\subsection{One-dimensional Image Model}
\begin{figure}
  \centering
  \includegraphics[scale=0.45]{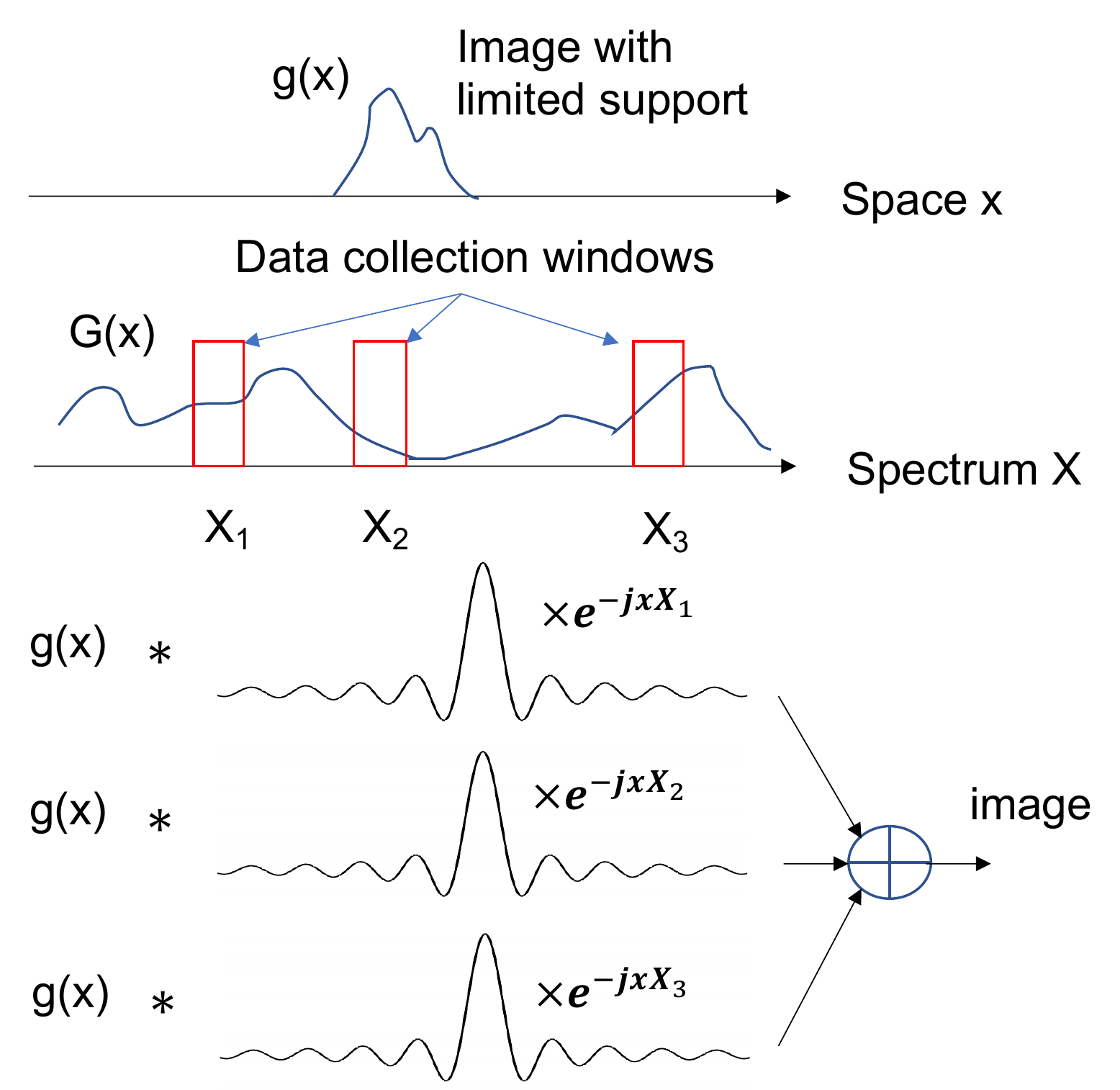}
  \caption{Reconstruction of image with multiple base stations.}\label{fig:recon}
\end{figure}

For simplicity of analysis and intuitive explanation, we adopt the 1-dimensional image model, which has been adopted in \cite{Munson1984}. As illustrated in Fig. \ref{fig:recon}, we consider a one-dimensional image centered at the origin. Then, $N$ base stations collect the data with windows centered at $X_1$, ..., $X_N$ and bandwidth $W$ (in Fig. \ref{fig:recon}, $N=3$). We simply consider all unknown spectrum data as zero, and then carry out inverse Fourier transform to reconstruct the original image. 

\subsection{Intuition}

The intuition of the image reconstruction is illustrated in Fig. \ref{fig:recon}. The data obtained by base station $n$ is given by
\begin{eqnarray}
G_n(X)=G(X)R(X-X_n),
\end{eqnarray}
where $R(X)$ is the unit rectangle function centered at 0. The reconstructed image is given by
\begin{eqnarray}\label{eq:reconstruct}
\hat{g}(x)&=&\mathcal{F}^{-1}\left(\sum_{n=1}^N G_n(X)\right)\nonumber\\
&=&\sum_{n=1}^N\mathcal{F}^{-1}\left(G(X)R(X-X_n)\right)\nonumber\\
&=&\sum_{n=1}^N g(x)\ast \left(\mbox{sinc}(x)e^{-jxX_n}\right)\nonumber\\
&=& g(x)\ast \left(\mbox{sinc}(x)\sum_{n=1}^Ne^{-jxX_n}\right)
\end{eqnarray}
where $\mathcal{F}$ means one-dimensional Fourier transform, the third equality is due to the duality of multiplication and convolution. From (\ref{eq:reconstruct}), we observe  the reconstructed image is the superposition of the reconstructions by each base station. The sinc function blurs each reconstructed images by each base station, due to its sidelobes. However, each sinc function is scaled by the phase factor $e^{-jxX_n}$, where $X_n$ is different for different base stations. The original image corresponds to $x=0$, where $e^{-jxX_n}=1$ and the different reconstructed images enhance each other. The interesting thing lies in the sidelobes of the sinc functions, where $x>0$. Due to the different $X_n$'s, different sinc functions, scaled by $e^{-jxX_n}$, will have different phases, thus may cancel each other. Therefore, when using the data from multiple base stations, the ambiguity due to the sidelobes of the sinc function will be statistically weakened, thus improving the quality of the reconstructed signal.

\subsection{Imaging Error Analysis}
For the error analysis of imaging, we assume a large number of base stations and model the sidelobe cancellation effect using a random process. To this end, we define $s(x)=\sum_{n=1}^Ne^{-jxX_n}$. Obviously, when $x=0$, we have $s(0)=N$. It is easy to show
\begin{eqnarray}
E[s(x)]=\sum_{n=1}^NE\left[e^{-jxX_n}\right]=0, \qquad \forall x\in \mathbb{R}.
\end{eqnarray}

Based on the following fact,
\begin{eqnarray}
Var\left[e^{-jxX_n}\right]=1,
\end{eqnarray}
according to the central limit theorem, we have
\begin{eqnarray}
\frac{1}{\sqrt{N}}\lim_{N\rightarrow\infty} s(x) \sim \mathcal{N}_c(0,1),
\end{eqnarray}
when $x\neq 0$, where $\mathcal{N}_c$ means circular symmetric Gaussian distribution. 

We further calculate the autocorrelation of $s(x)$, where $\delta x\neq 0$:
\begin{eqnarray}
&&E[s(x)s^*(x+\delta x)]\nonumber\\
&=&E\left[\sum_{n=1}^Ne^{-jxX_n}\sum_{m=1}^Ne^{jxX_m}\right]\nonumber\\
&=&\sum_{n=1}^N E \left[e^{j\delta xX_n}\right]
+\sum_{n\neq m} E \left[e^{-j(xX_n-xX_m-\delta x X_m)}\right]\nonumber\\
&=&\sum_{n\neq m}E \left[e^{-jxX_n}\right]E \left[e^{-j(x+\delta x) X_m}\right]\nonumber\\
&=&0,
\end{eqnarray}
where the third equality is due to the fact $E \left[e^{j\delta xX_n}\right]$, and factorization in the fourth equality is due to the independence of $\{X_n\}_{n=1,...,N}$. Therefore, we can model $s(x)$ as an i.i.d. Gaussian random sequence with variance $\frac{1}{N}$, except for $s(0)=N$.

The reconstruction of the image is then given by the following output of linear system:
\begin{eqnarray}
\hat{g}(x)=(g\ast h)(x),
\end{eqnarray}
where the impulse response $h$ is given by
\begin{eqnarray}
h(x)=\frac{s(x)\mbox{sinc}(x)}{\sqrt{\sum_x |s(x)|^2\mbox{sinc}^2(x)}}.
\end{eqnarray}
Here the denominator is to normalize energy of $h(x)$. Due to the law of large numbers, we have
\begin{eqnarray}
h(0)&=&\frac{N}{\sqrt{N^2+\frac{1}{N}\sum_{t\neq 0}\mbox{sinc}^2(t)}}\nonumber\\
&\approx&1-\frac{1}{2N^3}\sum_{t\neq 0}\mbox{sinc}^2(t)
\end{eqnarray}

We consider the MSE of the estimated complex reflectivity, which is given by
\begin{eqnarray}
E[|g(x)-\hat{g}(x)|^2]&=&E[|g\ast \delta (x)-g\ast h(x)|^2]\nonumber\\
&=&E[|g\ast \Delta h (x)|^2],
\end{eqnarray}
where $\delta$ is the discrete time delta function and $\delta h$ is the difference between $\Delta$ and $h$. Then, we further have 
\begin{eqnarray}
&&E[|g(x)-\hat{g}(x)|^2]\nonumber\\
&=&E\left[\sum_{t}g(t)\Delta h(x-t)\sum_{s}g^*(s)\Delta h^*(x-s)\right]\nonumber\\
&=&E\left[\sum_{t}|g(t)|^2|\Delta h(x-t)|^2\right]\nonumber\\
&=&|g(x)|^2|\Delta h(0)|^2+\sum_{t\neq 0}|g(t)|^2E[|\Delta h(x-t)|^2]\nonumber\\
&=&\frac{1}{2N^3}|g(x)|^2\sum_{t\neq 0}\mbox{sinc}^2(t)\nonumber\\
&+&\frac{1}{N}\sum_{t\neq x}|g(t)|^2\mbox{sinc}^2(x-t),
\end{eqnarray}
Therefore, the MSE decreases inversely proportional to $N$, when $N$ is sufficiently large.

\section{Numerical Results}\label{sec:numerical}

\begin{figure}
  \centering
  \includegraphics[scale=0.55]{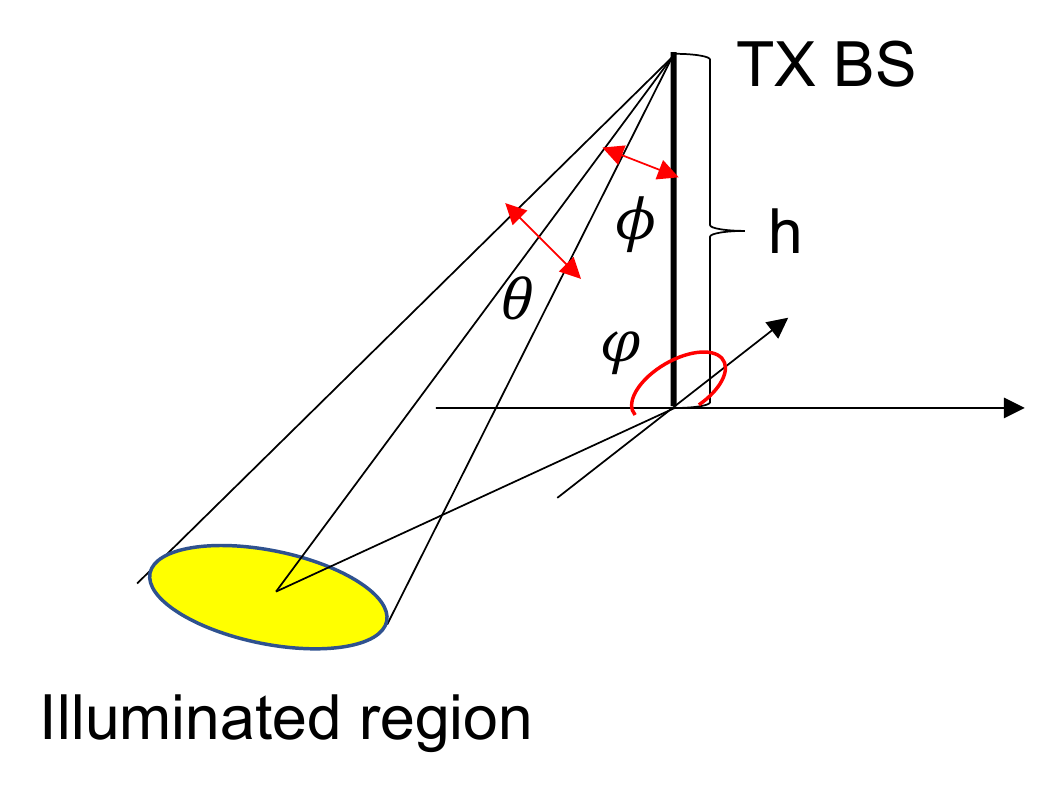}
  \caption{Geometry of base station, beam and illumiination.}\label{fig:geometry}
\end{figure}

We consider a region, in which the base stations are located on a regular grid. In each time slot, each base station sends out a beam which is modeled as a cone with open angle $\theta$ and tilt angle $\phi$, as illustrated in Fig. \ref{fig:geometry}. Then, the projection o the cone on the ground is an ellipse, whose center is given by
\begin{eqnarray}
\mathbf{c} = \mathbf{p}+h\tan(\phi)(\cos(\psi),\sin(\psi)),
\end{eqnarray}
where $\psi$ is the planar angle. The eccentricity of the ellipse is given by \cite{Thomas1979}
\begin{eqnarray}
e=\frac{\sin\phi}{\cos\frac{\theta}{2}},
\end{eqnarray}
where $\theta$ is the open angle of the beam. The major axis length of the ellipse is given by
\begin{eqnarray}
a=\frac{h}{2}\left(\tan\left(\phi+\frac{\theta}{2}\right)-\tan\left(\phi-\frac{\theta}{2}\right)\right).
\end{eqnarray}
The minor axis length is then obtained by
\begin{eqnarray}
b=a\sqrt{1-e^2}. 
\end{eqnarray}

\begin{figure}
  \centering
  \includegraphics[scale=0.45]{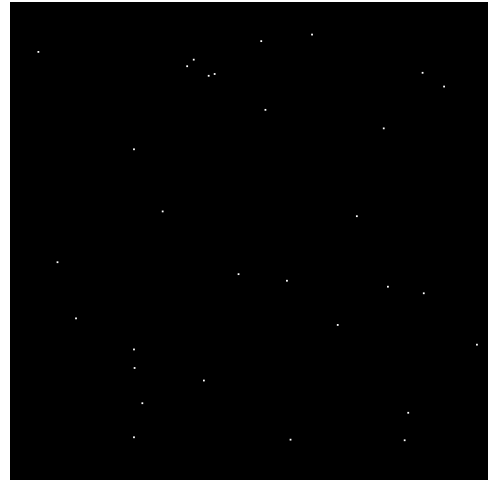}
  \caption{Original distributions of significant reflectors.}\label{fig:true}
\end{figure}

\begin{figure}
  \centering
  \includegraphics[scale=0.45]{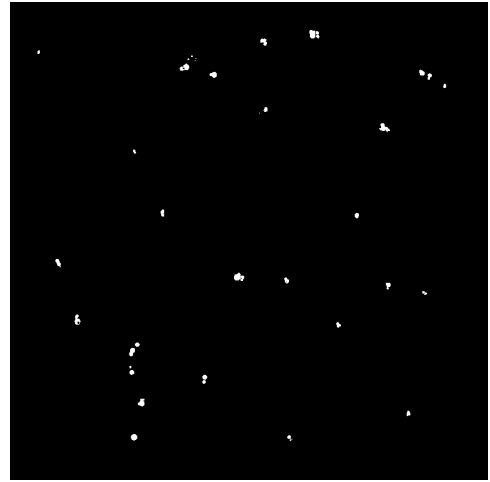}
  \caption{Reconstruction of significant reflectors.}\label{fig:recovery}
\end{figure}

We randomly place 30 significant reflectors in a 800m$\times$800m square region, each being a 3m$\times$3m square. The background is dark. We consider 256 subcarriers and the total bandwidth of signal is 512MHz, in the band of 5GHz. We assume that the base stations form a grid with spacing of 200m. The open angle of beam is 5 degrees. At each time slot, the probability for each base station to transmit is 0.05, which randomly selects one of 5 channels to transmit, while the receivers also randomly select the channels. We further assume that only base stations within 400 meters to the illuminated region receive the signal for further analysis, since a further distance will yield a substantially low signal power. For simplicity, we consider only the noise-free case, and will leave the impact of noise to the future research. This is reasonable since we assume a very small beam open angle, such that the power path loss is small. We also consider 64 antennas at each base stations, with the spacing of half carrier wavelength. Numerical results show that the cross direction resolution is very coarse. Therefore, we can only obtain a reasonable resolution in the radius direction. Hence, we use the fusion algorithm in Section \ref{sec:insuff}. The comparison between the original image and the reconstructed image is given in Figures \ref{fig:true} and \ref{fig:recovery}. We observe that most of the significant reflectors have been identified, while there are also some false detections. 

\section{Conclusion}\label{sec:conclusion}
In this paper, we have studied the imaging of outdoor environment using communication signals and infrastructure. The imaging principles of SAR and ISARs are leveraged for the communication signal based imaging, by addressing the difference of the scenarios. The framework of Fourier transform is employed for the image reconstruction. Various challenges, such as distributed data patches, missing data and 3D imaging, have been addressed. Numerical simulations have been carried out for the imaging of significant reflectors, which has demonstrated the validity of the proposed algorithms. 

\appendices

\section{Tradeoff Between Communications and Sensing}
The capability of sensing the environment using communication signals stems from the impact of environment variety on the communications. Consequently, there is a tradeoff between communications and sensing: when the communication signal is more sensitive to the environment impact, more information on the environment can be gleaned, which meanwhile incurs more performance degradation for communications. Therefore, it is important to explore the tradeoff between sensing and communications. 

\begin{figure}
  \centering
  \includegraphics[scale=0.35]{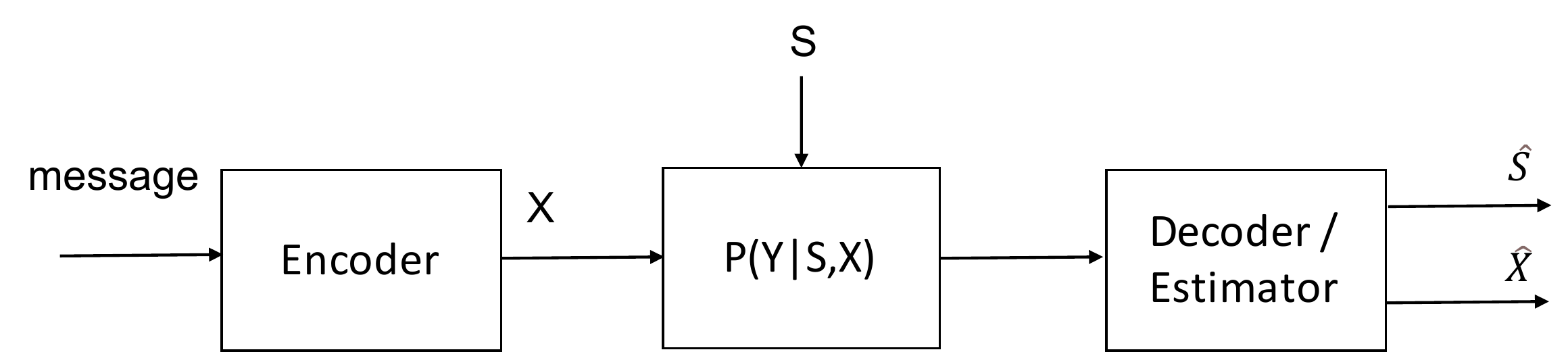}
  \caption{Model of one-way joint communications and sensing.}\label{fig:forward}
\end{figure}

We consider the model in Fig. \ref{fig:forward}. where the communication signal and environment are denoted by $X$ and $S$, and the channel is characterized by the conditional probability $P(Y|S,X)$.
Using the simple information equalities, we have
\begin{eqnarray}
I(X;Y)&=&H(Y)-H(Y|X)\nonumber\\
&=&H(Y)-(H(Y|X,S)+I(Y;S|X)),
\end{eqnarray}
which results in
\begin{eqnarray}\label{eq:tradeoff}
I(X;Y)+I(Y;S|X)=H(Y)-H(Y|X,S).
\end{eqnarray}
Although the above derivation is straightforward, the information theoretic meaning is interesting:
\begin{itemize}
\item The terms $I(X;Y)$ and $I(Y;S|X)$ indicate the information that can be retrieved for the communication information and channel information, respectively, from the receiver measurements. Notice that we use $I(Y;S|X)$, instead of $I(Y;S)$, based on the assumption that the receiver can reliably decode the transmitted codewords and recover $X$. Therefore, the left hand side is the sum of the information on the communication message and the channel.

\item The right hand side is the total uncertainty in the received signal $Y$, excluding the term $H(Y|X,S)$ that indicates the randomness incurred by the channel noise. Hence, this difference represents the effective information in the signal $Y$. 
\end{itemize}

In summary, the tradeoff between communications and sensing is characterized by the upper bound for the sum of the information in bits. When more information can be derived from the measurement, less bits of communication message can go through the channel, thus impairing the communication performance. 

Now, we assume that there exists a deterministic function $g$ such that
\begin{eqnarray}
Y(t)=g(X(t),S)+N(t), \qquad t=1,...M,
\end{eqnarray}
where $N(t)$ is Gaussian white noise with power $\sigma_n^2$. Then, if the received signal power is $P$, the tradeoff in (\ref{eq:tradeoff}) becomes the following inequality:
\begin{eqnarray}
I(X;Y)+I(Y;S|X)\leq \frac{1}{2}\log_2\left(1+\frac{P}{\sigma_n^2}\right),
\end{eqnarray}
where the inequality stems from the fact that the Gaussian distribution achieves the maximum entropy.

\end{document}